\documentclass[journal]{IEEEtran}

\usepackage{times, amsmath, epsfig, amssymb}
\usepackage{multirow}
\usepackage{stfloats}
\usepackage{cite}
\usepackage{tabu}
\usepackage{multicol}
\usepackage{multirow}
\usepackage{subdepth}
\usepackage{mathtools}
\usepackage{epstopdf}
\usepackage{microtype}
\usepackage{xcolor}
\usepackage{diagbox}
\usepackage[font=footnotesize]{caption}
\usepackage{array}
\usepackage[font=footnotesize]{subcaption}
\usepackage[hidelinks]{hyperref}

\newcommand\tstruta{\rule{0pt}{10pt}}
\newcommand\tstrutb{\rule{0pt}{10pt}}
\newcommand\tstrutc{\rule{0pt}{10pt}}

\newcommand{\figurewidth}{0.96\columnwidth}
\newcommand{\comma}{\text{,}}
\newcommand{\TT}{^\mathrm{T}}
\newcommand{\HH}{^\mathrm{H}}
\newcommand{\E}{\mathbb{E}}
\DeclareMathOperator*{\lgg}{\text{log}_2}

\title{Analysis and Augmented Spatial Processing for Uplink OFDMA MU-MIMO Receiver with Transceiver I/Q Imbalance and External Interference}

\begin{document}


\author{Aki Hakkarainen, \emph{Student Member, IEEE}, Janis Werner,\\Kapil R. Dandekar, \emph{Senior Member, IEEE}, and Mikko Valkama,
\emph{Senior Member, IEEE}}


\maketitle


\let\thefootnote\relax\footnotetext{
\footnotesize
This work has been submitted to the IEEE for possible publication. Copyright may be transferred without notice, after which this version may no longer be accessible.

This work was supported in part by the Finnish Funding Agency for Technology and Innovation (Tekes) under the projects “Reconﬁgurable Antenna-based Enhancement of Dynamic Spectrum Access Algorithms” and “Future Uncoordinated Small-Cell Networks Using Reconﬁgurable Antennas,” in part by the Industrial Research Fund of Tampere University of Technology, in part by the Academy of Finland under the projects 251138, 284694 and 288670, in part by the Doctoral Programme of the President of Tampere University of Technology, in part by the Foundation of Nokia Corporation, and also in part by the National Science Foundation (NSF) under Grant CNS-1457306. Preliminary work addressing a limited subset of initial results was presented at the IEEE Global Communications Conference (GLOBECOM), Austin, TX, USA, December 2014~\cite{hakkarainen_precoded_2014}, and at International Symposium on Wireless Communication Systems (ISWCS), Brussels, Belgium, August 2015~\cite{hakkarainen_transceiver_2015}. 

A. Hakkarainen, J. Werner, and M. Valkama are with the Department of Electronics and Communications Engineering, Tampere University of Technology, Tampere 33720, Finland (e-mail: aki.hakkarainen@tut.fi janis.werner@tut.fi; mikko.e.valkama@tut.fi). 

K. R. Dandekar is with the Department of Electrical and Computer Engineering, Drexel University, Philadelphia, PA 19104 USA (e-mail: \mbox{dandekar@coe.drexel.edu}).
}


\begin{abstract} 

This paper addresses receiver (RX) signal processing in multiuser multiple-input multiple-output (MU-MIMO) systems. We focus on uplink orthogonal frequency-division multiple access (OFDMA)-based MU-MIMO communications under in-phase/quadrature (I/Q) imbalance in the associated radio frequency electronics. It is shown in the existing literature that transceiver I/Q imbalances cause cross-talk of mirror-subcarriers in OFDM systems. As opposed to typically reported single-user studies, we extend the studies to OFDMA-based MU-MIMO communications, with simultaneous user multiplexing in both frequency and spatial domains, and incorporate also external interference from multiple sources at RX input, for modeling challenging conditions in increasingly popular heterogeneous networks. In the signal processing developments, we exploit the augmented subcarrier processing, which processes each subcarrier jointly with its counterpart at the image subcarrier, and jointly across all RX antennas. Furthermore, we derive an optimal augmented linear RX in terms of minimizing the mean-squared error. The novel approach integrates the I/Q imbalance mitigation, external interference suppression and data stream separation of multiple UEs into a single processing stage, thus avoiding separate transceiver calibration. Extensive analysis and numerical results show the signal-to-interference-plus-noise ratio (SINR) and symbol-error rate (SER) behavior of an arbitrary data stream after RX spatial processing as a function of different system and impairment parameters. Based on the results, the performance of the conventional per-subcarrier processing is heavily limited under transceiver I/Q imbalances, and is particularly sensitive to external interferers, whereas the proposed augmented subcarrier processing provides a high-performance signal processing solution being able to detect the signals of different users as well as suppress the external interference efficiently. Finally, we also extend the studies to massive MIMO framework, with very large antenna systems. It is shown that, despite the huge number of RX antennas, the conventional linear processing methods still suffer heavily from I/Q imbalances while the augmented approach does not have such limitations.
\end{abstract}



\section{Introduction} 

Modern communication systems need to support the ever-increasing user needs of faster data connections and cheaper devices. This has resulted, e.g., in adopting larger and more complicated symbol alphabets which are, unfortunately, also more vulnerable to various signal distortions than conventional solutions. In addition, the user equipment (UE), including also the analog radio frequency (RF) circuitry, should be implemented with very low costs and silicon area. These things, among other requirements of maximum performance, low power, small size etc., have resulted in a situation where the RF imperfections and their mitigation methods by cost-efficient digital signal processing have become very important aspects in system design. One of these RF imperfections is the so-called in-phase/quadrature (I/Q) imbalance which occurs in direct-conversion transceivers~\cite{mirabbasi_classical_2000}. Physically, when the baseband signal is up-converted in the transmitters (TXs) or when the RF signal is down-converted in the receivers (RXs), the signals in the I and Q branches have slight differences in their amplitude and phase responses, e.g., due to manufacturing tolerances. This leads to imbalance between the I and Q signals and thus distorts the overall signal waveforms~\cite{schenk_rf_2008}.  

I/Q imbalance effects and mitigation are widely studied for orthogonal frequency division multiplexing (OFDM) waveforms. In~\cite{tarighat_compensation_2005, tarighat_joint_2007, ozdemir_exact_2013, ozdemir_exact_2014, krone_capacity_2008, narasimhan_digital_2009}, I/Q imbalance in single-input single-output (SISO) OFDM systems is studied comprehensively. The SISO approach is extended to cover multiple TX antennas in~\cite{maham_impact_2012, ozdemir_sinr_2012} while~\cite{tarighat_mimo_2005, schenk_estimation_2006, schenk_performance_2007, schenk_rf_2008, ozdemir_i/q_2013} consider multiple antennas on both TX and RX sides, resulting in full multiple-input multiple-output (MIMO) communications in single-user context (i.e., SU-MIMO). The joint effects of I/Q imbalance and power amplifier nonlinearities are studied in~\cite{qi_compensation_2010} whereas~\cite{tandur_joint_2007} focuses on I/Q imbalance with carrier frequency offset and \cite{tubbax_compensation_2005, zou_joint_2009} consider I/Q imbalance with phase noise, all again in single-user context. Based on the studies listed above, the so-called augmented subcarrier processing for I/Q imbalance mitigation in OFDM systems has been proposed in~\cite{tarighat_mimo_2005, tarighat_compensation_2005, schenk_estimation_2006, tarighat_joint_2007, narasimhan_digital_2009, ozdemir_i/q_2013, ishaque_capacity_2013, yoshida_analysis_2009}. Therein, each subcarrier signal is processed jointly with the corresponding signal at the image, or mirror, subcarrier. This approach is very close to widely-linear processing~\cite{picinbono_widely_1995} where the signal and its complex conjugate are processed jointly. The widely-linear processing is originally proposed for processing non-circular signals, see e.g.~\cite{picinbono_widely_1995, schreier_second-order_2003, adali_complex-valued_2011}, and also for time-domain I/Q imbalance mitigation, see e.g.~\cite{valkama_blind_2005, anttila_circularity-based_2008}, since I/Q imbalance results in non-circular signals even with originally circular signals. 

Although the single-user OFDM studies listed above concentrate on the I/Q imbalance challenges and their mitigation methods, they do not address multiuser MIMO (MU-MIMO)~\cite{gesbert_shifting_2007} aspects, i.e., having multiple UEs transmitting simultaneously at a given subcarrier. Furthermore, the above works do not address UE multiplexing in frequency domain, through orthogonal frequency division multiple access (OFDMA) principle, either. UE multiplexing through single-carrier frequency-division multiple access (SC-FDMA) together with joint TX+RX I/Q imbalances is studied in \cite{gomaa_multi-user_2011, ishaque_capacity_2013} while TX I/Q imbalance with SC-FDMA and OFDMA is studied in \cite{yoshida_analysis_2009}. However, these studies consider a case where each subcarrier is allocated only to a single single-antenna UE which transmits towards a single-antenna BS. Furthermore, none of the studies listed above take the influence of possible external interferers into account. Some of the studies also make somewhat limited assumptions of equal I/Q imbalance coefficients between different subcarriers and/or transceiver branches. Our earlier study in~\cite{hakkarainen_interference_2014} focused on the external interference suppression with antenna array processing in OFDM systems but the study was limited only to the single-user single-input multiple-output (SU-SIMO) scenario. The rather limited SU-SISO and SU-MIMO schemes are considerable simpler than the full MU-MIMO transmission from the viewpoints of the signal models and associated signal processing algorithms. Therefore, in this paper, we extend the existing results towards more generic MU-MIMO systems, incorporating also the large antenna system or massive MIMO~\cite{hoydis_massive_2013, larsson_massive_2014, lu_overview_2014} aspects, receiving increasing interest currently. In particular, {\it the main contributions of this paper are the following}:
\begin{itemize}
	\item In the analysis and mitigation, we focus on a generic uplink MU-MIMO OFDMA system under transceiver I/Q imbalances. This means that {\it multiple UEs  transmit simultaneously towards the BS at each of the available subcarriers}, and that further UE multiplexing takes place simultaneously in the frequency domain. Such a multiple-access scheme is already adopted to IEEE802.16 Broadband Wireless Metropolitan Area Networks (WiMAX) advanced air interface specification~\cite{wimax_2012} and has been considered to be a potential air interface technology for the future wireless local area network (WLAN) implementations within the IEEE 802.11ax/HEW framework~\cite{ieee80211sg, ieee80211tg}. In addition, the considered model can be easily applied to other multicarrier systems such as 3GPP long term evolution (LTE) and LTE-Advanced which, in terms of uplink, are based on SC-FDMA waveform.

	\item We also include the effects of external interferers into the analysis and show how antenna array processing can be efficiently used to suppress the external interference having a given spatial response, in spite of I/Q imbalances. This kind of external interferers may exist, e.g., in increasingly popular heterogeneous networks where the UEs at the cell-edge of a macro cell, and consequently with considerably high TX power levels, severely interfere with the reception in a co-channel neighboring femto-cell BS.

	\item We formulate our analysis in a generic and flexible way by allowing arbitrary system parameters. This approach allows us to use frequency-selective and transceiver branch-dependent I/Q imbalance parameters in the analysis and signal processing. 

	\item The developed augmented subcarrier processing introduces a novel combining approach which can jointly separate all spatially multiplexed UE data streams from each others as well as mitigate the effects of I/Q imbalances and external interference, thus avoiding separate transceiver calibration.

	\item We also extend the studies to massive MIMO framework, i.e., to cases where the number of RX antennas is an order of magnitude higher than the number of spatially multiplexed users, and show the sensitivity of such systems to transceiver I/Q imbalances with different RX spatial processing schemes.

	\item Finally, we provide an extensive set of numerical experiments which illustrate explicitly the influence of different system parameters under the inevitable RF imperfections. 
\end{itemize}
With these considerations we can provide valuable insight for future MU-MIMO OFDMA system designers as well as a fundamental starting point for future research. One of the central technical findings is that the performance of conventional per-subcarrier spatial processing is heavily limited under transceiver I/Q imbalances, and is particularly sensitive to external interferers, whereas the proposed augmented spatial subcarrier processing provides a robust and high-performance RX signal processing solution being able to detect the data streams of different users as well as suppress the effects of the external interference in a highly efficient manner, in spite of transceiver I/Q imbalances. Another central finding is that massive MIMO systems can, indeed, be sensitive to RF chain I/Q imbalances, in spite of high processing gain stemming from the massive number of antenna units. This is an outcome that more simplified modeling based studies reported, e.g., in~\cite{bjornson_massive_HW_2014, bjornson_massive_2014}, have not clearly reported. 

The rest of the paper is organized as follows. Section II presents the fundamental MU-MIMO OFDMA signal and system models under transceiver I/Q imbalances. Linear minimum mean-square error (LMMSE) and augmented LMMSE RXs are derived in Section III along with output signal-to-interference-plus-noise ratio (SINR) and computational complexity analyses. Section IV gives extensive numerical evaluations and illustrations as a function of numerous system parameters. Finally, we conclude the paper in Section V. 

\emph{Notation:} Throughout this paper, vectors and matrices are written with bold characters. The superscripts $(\cdot)\TT$, $(\cdot)\HH$, $(\cdot)^{\ast}$ and $(\cdot)^{-1}$ represent transpose, Hermitian (conjugate) transpose, complex conjugate and matrix inverse, respectively. The tilde sign $\tilde{({\cdot})}$ is used to present an augmented quantity and the results obtained by the augmented processing. 
We write $\text{diag} \left(x_{11}, x_{22}, \cdots , x_{ii} , \cdots \right)$ to denote a diagonal matrix $\mathbf{X}$ that is composed of the entries $x_{ii}$ on the main diagonal. The natural basis vector, where the $q^{\text{th}}$ entry is equal to one and the rest are zeros, is denoted as $\mathbf{e}_q$. The statistical expectation is denoted with $\E [ \cdot ]$. A complex random variable $x$ is called circular if $\E [ x^2 ] = 0$.

\section{Fundamental Signal and System Models}

\begin{figure}[t]
	\centering
	\includegraphics[width=\figurewidth]{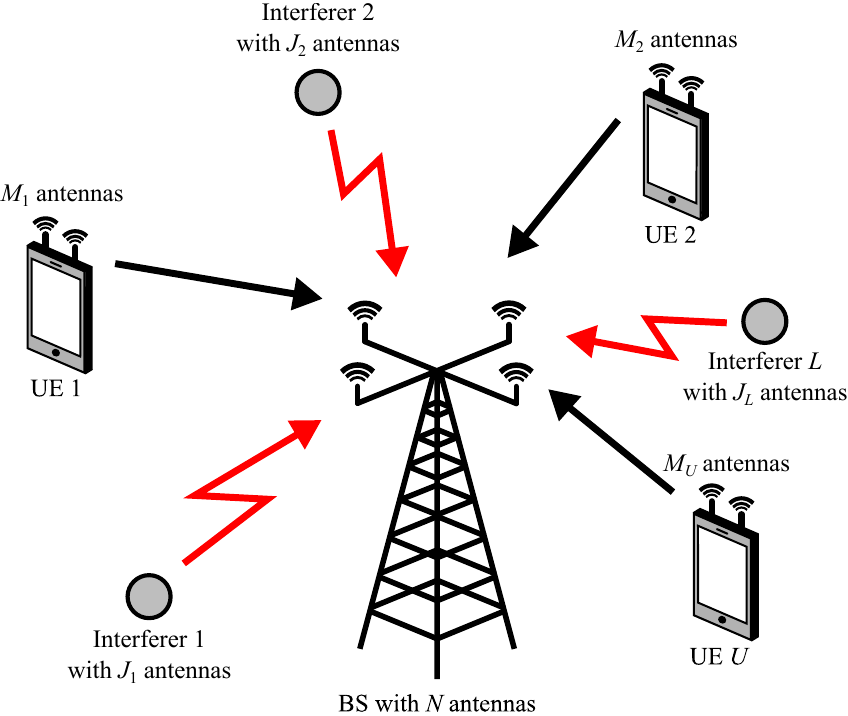}
	\vspace{6pt}
	\caption{A general MU-MIMO uplink scenario with a single base station, $U$ mobile users and $L$ external interferers, all being simultaneously active at subcarrier $c$. The base station is equipped with $N$ RX antennas whereas UE $u$ has $M_u$ and interferer $l$ has $J_l$ TX antennas. Further user multiplexing takes place in the frequency domain, through the OFDMA principle.}
	\label{MU-MIMO_setup}
\end{figure}

OFDM and OFDMA systems are based on multicarrier transmission where the parallel subcarriers are modulated and deployed independently and where users can be flexibly multiplexed in both frequency and spatial domains. On the TX side, multiple parallel frequency domain data streams are jointly converted to time domain through the inverse fast Fourier transform (IFFT). On the RX side, the received time domain signal is then converted back to frequency domain data streams through the fast Fourier transform (FFT). Since the essential data exist at the subcarrier level, we analyze an uplink OFDMA MU-MIMO system from an arbitrary subcarrier point of view. The generic uplink system model comprises a single BS which serves multiple UEs simultaneously at each subcarrier. The subcarriers are indexed with \mbox{$c \in \{-C/2, \dotsc , -1, 1, \dotsc , C/2\}$} where $C$ is the total number of active subcarriers. Additionally, the image (or mirror) subcarrier is defined as $c' = -c$. The number of UEs spatially multiplexed at subcarrier $c$ is denoted with $U$ while the corresponding number at subcarrier $c'$ is $V$. Correspondingly, the users are indexed by \mbox{$u \in \{ 1, \dotsc , U \}$} and \mbox{$v \in \{ 1, \dotsc , V \}$}. Note that depending on the subcarrier allocation for the UEs, $u$ and $v$ might sometimes refer to the same UE if it is transmitting at both subcarriers $c$ and $c'$. The BS has $N$ RX antennas whereas UE $u$ is equipped with $M_u$ TX antennas. In addition, the effect of $L$ external interferers is included to the model and external interferer $l$ is assumed to have $J_l$ TX antennas. The scenario under consideration is illustrated in Fig.~\ref{MU-MIMO_setup}. 

We denote the transmitted baseband equivalent spatial signal vector of user $u$ at subcarrier $c$ by \mbox{$\mathbf{s}_{u \comma c} = \mathbf{G}_{u \comma c} \mathbf{x}_{u \comma c} \in {\mathbb{C}}^{M_u \times 1}$}. Here, $\mathbf{x}_{u \comma c} \in {\mathbb{C}}^{Q_{u \comma c} \times 1}$ denotes the parallel transmitted data streams of UE $u$ at subcarrier $c$ and $Q_{u \comma c}$ is the number of these streams. In addition, $\mathbf{G}_{u \comma c} \in {\mathbb{C}}^{M_u \times Q_{u \comma c}}$ denotes the precoder matrix which maps the actual data snapshots to TX antennas. Similarly, the transmitted baseband equivalent signal snapshot vector of user $v$ at the image subcarrier $c'$ is given by \mbox{$\mathbf{s}_{v \comma c'} = \mathbf{G}_{v \comma c'} \mathbf{x}_{v \comma c'} \in {\mathbb{C}}^{M_v \times 1}$}. 
Finally, $s_{\text{int,} l \comma c} \in{\mathbb{C}}^{J_l \times 1}$ denotes the signal snapshot vector originating from the $l^{\text{th}}$ external interferer at subcarrier $c$. Throughout this paper, we assume that in all associated devices each antenna is connected to a separate transceiver chain. All signal vectors refer to subcarrier-level (frequency-domain) quantities in the considered OFDMA radio system, i.e., before IFFT in the TXs and after FFT in the RXs. The most essential variables used throughout the paper are listed in Table \ref{variable_table}. 

\subsection{TX and RX I/Q Imbalance Characteristics}

The imperfections in the analog electronics of direct-conversion transceivers create I/Q imbalance~\cite{mirabbasi_classical_2000}. On the one hand, the gain imbalance $g$ is created by unequal gains or attenuations between the I and Q branches in amplifiers, filters, mixers and digital-to-analog and analog-to-digital converters. On the other hand, the phase imbalance $\phi$ occurs mainly due to the imperfections in mixers and phase shifters, as well as due to phase response differences of the branch filters. In general, both the gain and phase imbalance are \emph{frequency-dependent} already within a few MHz processing bandwidths~\cite{anttila_frequency-selective_2008, luo_digital_2011}, and thus need to be modeled accordingly.

\begin{table*}[t]
	\center
	\caption{Most important variables used throughout the paper}
	\resizebox{\textwidth}{!}{
	\begin{tabular}{ l | l | l }
		\hline \tstruta
		Variable	 &Dimensions &Definition
		\\ \hline \hline \tstrutb
		$\sigma_{n \comma c}^2$	&scalar		&Noise power at subcarrier $c$ \\
		$\sigma_{x \comma u \comma c}^2$, $\sigma_{x \comma v \comma c'}^2$	&scalars	&Power of a single data stream of UE $u$ at subcarrier $c$ and of UE $v$ at subcarrier $c'$ \\
		$J_l$	&scalar		&Number of TX antennas of external interferer $l$ \\
		$L$		&scalar		&Number of external interferers \\
		$M_u$	&scalar		&Number of TX antennas of UE $u$\\
		$N$		&scalar		&Number of BS RX antennas \\
		$Q_{u \comma c}$, $Q_{v \comma c'}$	&scalars	&Number of parallel transmitted data streams by UE $u$ at subcarrier $c$ and by UE $v$ at subcarrier $c'$\\
		$S$		&scalar		&Total number of all transmitted data streams from all UEs at subcarrier $c$, i.e., $S = \sum_{u=1}^U Q_{u \comma c}$ \\
		$U, V$ 	&scalars		&Number of spatially multiplexed UEs at subcarriers $c$ and $c'$\\
		$c$ 		&scalar 		&Subcarrier index \\
		$c'$		&scalar		&Image subcarrier index \\
		$u, v$	&scalars		&UE indeces for subcarriers $c$ and $c'$\\
		$\widetilde{\mathbf{\Psi}}_{u \comma c}$, $\widetilde{\mathbf{\Omega}}_{u \comma c}$	&$N \times M_u$	&Total effective channel matrices including the joint effects of TX+RX I/Q imb.\ as well as the wireless channel \\
		$\mathbf{G}_{u \comma c}$, $\mathbf{G}_{v \comma c'}$	&$M_u \times Q_{u \comma c}$, $M_v \times Q_{v \comma c'}$\!\!\! & Precoder matrices for UE $u$ at subcarrier $c$ and for UE $v$ at subcarrier $c'$\\
		$\mathbf{H}_{u \comma c}$, $\mathbf{H}_{u \comma c'}$	&$N \times M_u$ 	&Channel response matrices of UE $u$ at subcarriers $c$ and $c'$\\
		$\mathbf{K}_{\text{Rx1,} c}$, $\mathbf{K}_{\text{Rx2,} c}$		&$N \times N$	&Diagonal RX I/Q imbalance matrices of BS at subcarrier $c$ \\
		$\widetilde{\mathbf{K}}_{\text{RxA,} c}$, $\widetilde{\mathbf{K}}_{\text{RxB,} c}$	&$2N \times N$ 	&Augmented RX I/Q imbalance matrices at subcarrier $c$ \\
		$\mathbf{K}_{\text{Tx1,} u \comma c}$, $\mathbf{K}_{\text{Tx2,} u \comma c}$ 	&$M_u \times M_u$	&Diagonal TX I/Q imbalance matrices of UE $u$ at subcarrier $c$ \\
		$\widetilde{\mathbf{R}}_{\text{TxRxi,} c}$	&$2N \times 2N$	&Covariance matrix of the augmented received signal vector under joint TX+RX I/Q imbalances at subcarrier $c$ \\
		$\mathbf{R}_{\text{z,} c}$, $\mathbf{R}_{\text{z,} c'}$		&$N \times N$		&Covariance matrices of external interference and noise at subcarriers $c$ and $c'$\\
		$\mathbf{W}_c$, $\widetilde{\mathbf{W}}_c$		&$N \times S$, $2N \times S$		&Combiner weighting matrices at subcarrier $c$ \\	
		$\mathbf{n}_c$				&$N \times 1$		&Additive noise in the RX electronics at subcarrier $c$ \\
		$\mathbf{s}_{\text{int,} l \comma c}$	&$J_l \times 1$	&Transmitted baseband equivalent spatial signal vector of interferer $l$ at subcarrier $c$ \\
		$\mathbf{s}_{u \comma c}$, $\mathbf{s}_{v \comma c'}$	&$M_u \times 1$, $M_v \times 1$	&Transmitted baseband equivalent spatial signal vector of UE $u$ at subcarrier $c$ and of UE $v$ at subcarrier $c'$\\
		$\mathbf{r}_{\text{TxRxi,} c}$, $\widetilde{\mathbf{r}}_{\text{TxRxi,} c}$	&$N \times 1$, $2N \times 1$	&Received signal vectors under joint TX+RX I/Q imbalances at subcarrier $c$ \\
		$\mathbf{v}_{\text{TxRxi,} q \comma u \comma c}$, $\widetilde{\mathbf{v}}_{\text{TxRxi,} q \comma u \comma c}$		&$N \times 1$, $2N \times 1$ 	&Cross-correlation vectors between the received signal vector and data stream $q$ of UE $u$ at subcarrier $c$ \\
		$\mathbf{x}_{u \comma c}$, $\mathbf{x}_{v \comma c'}$	&$Q_{u \comma c} \times 1$, $Q_{v \comma c'} \times 1$	&Transmitted data stream vectors of UE $u$ at subcarrier $c$ and of UE $v$ at subcarrier $c'$\\
		$\mathbf{y}_{\text{TxRxi,} c}$, $\widetilde{\mathbf{y}}_{\text{TxRxi,} c}$		&$S \times 1$	&Output signal vectors of the combiners under joint TX+RX I/Q imbalances at subcarrier $c$ \\
		$\mathbf{z}_c$, $\mathbf{z}_{c'}$		&$N \times 1$		&Sum of external interference and additive noise vectors at subcarriers $c$ and $c'$\\
		\hline
		\multicolumn{3}{l}
		{\multirow{2}{*}{The tilde sign $\tilde{({\cdot})}$ refers to augmented quantities and the results obtained by the augmented processing.}}
	\end{tabular} 
	}
	\label{variable_table}
\end{table*}

For notational convenience, we first define TX I/Q imbalance parameters  for a single TX antenna branch $m$ of user $u$ at subcarrier $c$. They are equal to $K_{\text{Tx1,} m \comma u \comma c} = (1+g_{\text{Tx,} m \comma u \comma c} e^{j\phi_{\text{Tx,} m \comma u \comma c}})/2$ and $K_{\text{Tx2,} m \comma u \comma c} = (1-g_{\text{Tx,} m \comma u \comma c} e^{j\phi_{\text{Tx,} m \comma u \comma c}})/2$ where $g_{\text{Tx,} m \comma u \comma c}$ and $\phi_{\text{Tx,} m \comma u \comma c}$ are the gain and phase imbalance coefficients for TX antenna branch $m$ of user $u$ at subcarrier $c$, respectively~\cite{schenk_rf_2008}. Since the UE has $M_u$ antennas and associated TX branches, we stack the I/Q imbalance parameters of different TX branches into diagonal matrices. Consequently, the TX I/Q imbalance matrices $\mathbf{K}_{\text{Tx1,} u \comma c}$ and $\mathbf{K}_{\text{Tx2,} u \comma c}$, both $\in {\mathbb{C}}^{M_u \times M_u}$, are given by 
\begin{align}
	\begin{split}
	&\mathbf{K}_{\text{Tx1,} u \comma c} = \text{diag}(K_{\text{Tx1,1,} u \comma c}, \cdots, K_{\text{Tx1,}M_{u} \comma u \comma c}), \\
	&\mathbf{K}_{\text{Tx2,} u \comma c} = \text{diag}(K_{\text{Tx2,1,} u \comma c}, \cdots, K_{\text{Tx2,}M_{u} \comma u \comma c}).
	\end{split}
\end{align}
Similarly, the I/Q imbalance characteristics for a single RX antenna branch $n$ at subcarrier $c$ are equal to $K_{\text{Rx1,} n, c} = (1+g_{\text{Rx,} n, c} e^{-j\phi_{\text{Rx,} n, c}})/2$ and $K_{\text{Rx2,} n, c} = (1-g_{\text{Rx,} n, c} e^{j\phi_{\text{Rx,} n, c}})/2$ where $g_{\text{Rx,} n, c}$ and $\phi_{\text{Rx,} n, c}$ denote the gain and phase imbalance coefficients of RX antenna branch $n$~\cite{schenk_rf_2008}. We stack also the RX I/Q imbalance parameters into diagonal matrices, resulting in the RX I/Q imbalance matrices $\mathbf{K}_{\text{Rx1,} c}$ and $\mathbf{K}_{\text{Rx2,} c}$, both $\in {\mathbb{C}}^{N \times N}$, given by
\begin{align}
	\begin{split}
	&\mathbf{K}_{\text{Rx1,} c} = \text{diag}(K_{\text{Rx1,1,} c}, \cdots, K_{\text{Rx1,}N, c}), \\
	&\mathbf{K}_{\text{Rx2,} c} = \text{diag}(K_{\text{Rx2,1,} c}, \cdots, K_{\text{Rx2,}N, c}).
	\end{split}
\end{align}
These matrices are used in the modeling  and analysis of the total effects of TX and RX imbalances in the considered MU-MIMO system. The above characterization allows setting the I/Q imbalance parameters freely and independently, not only between different UEs but also between different antenna branches of a single device. In addition, we assume that I/Q imbalance is frequency selective, i.e., I/Q imbalance parameters at different subcarriers are different. However, all derived expressions are valid also for the case where the I/Q imbalance parameters are equal in all transceivers, transceiver branches and/or subcarriers. 

\subsection{Uplink MU-MIMO Transmission under I/Q Imbalance}

The transmitted baseband equivalent signal snapshot vector of user $u$ at subcarrier $c$ under TX I/Q imbalance can be now written with the help of the TX I/Q imbalance matrices directly as~\cite{tandur_joint_2007}
\begin{align}
	\begin{split}
	\mathbf{s}_{\text{Txi,} u \comma c} &= \mathbf{K}_{\text{Tx1,} u \comma c} \mathbf{s}_{u \comma c}  + \mathbf{K}_{\text{Tx2,} u \comma c} \mathbf{s}_{u \comma c'}^*
	\\
	&= \mathbf{K}_{\text{Tx1,} u \comma c} \mathbf{G}_{u \comma c} \mathbf{x}_{u \comma c}  + \mathbf{K}_{\text{Tx2,} u \comma c} \mathbf{G}_{u \comma c'}^* \mathbf{x}_{u \comma c'}^*.
	\end{split}
	\label{s_Txi}
\end{align}%
Clearly, the structure of the transmitted signal is distorted, resulting in general in cross-talk between image-subcarriers $c$ and $c'$. This is already a well-established phenomenon in the existing literature, see e.g.~\cite{tandur_joint_2007, schenk_rf_2008, luo_digital_2011, ozdemir_exact_2013}. Notice, however, that if the image subcarrier $c'$ is not allocated for UE $u$ there is no cross-talk between the subcarriers of an individual UE and the resulting transmitted signal at subcarrier $c$ consists only of the scaled version of $\mathbf{s}_{u \comma c}$. 
However, when subcarrier $c'$ is allocated to another UE $v$, through the OFDMA principle, the corresponding emitted signal snapshot vector at subcarrier $c$ is of the form $\mathbf{s}_{\text{Txi,} v \comma c} = \mathbf{K}_{\text{Tx2,} v \comma c} \mathbf{s}_{v \comma c'}^* = \mathbf{K}_{\text{Tx2,} v \comma c} \mathbf{G}_{v \comma c'}^* \mathbf{x}_{v \comma c'}^*$. Then, when interpreted from RX perspective, this implies \emph{cross-talk or interference between UEs}. This issue is not addressed or taken into account in the existing literature~\cite{tandur_joint_2007, schenk_performance_2007, schenk_estimation_2006, tarighat_joint_2007, narasimhan_digital_2009, ozdemir_exact_2013, ozdemir_exact_2014, ozdemir_sinr_2012, maham_impact_2012, ozdemir_i/q_2013}. The corresponding transmitted signal vectors at the image subcarrier $c'$ are given by $\mathbf{s}_{\text{Txi,} u \comma c'} = \mathbf{K}_{\text{Tx1,} u \comma c'} \mathbf{G}_{u \comma c'} \mathbf{x}_{u \comma c'} + \mathbf{K}_{\text{Tx2,} u \comma c'} \mathbf{G}_{u \comma c}^* \mathbf{x}_{u \comma c}^*$ and $\mathbf{s}_{\text{Txi,} v \comma c'} = \mathbf{K}_{\text{Tx1,} v \comma c'} \mathbf{G}_{v \comma c'} \mathbf{x}_{v \comma c'} + \mathbf{K}_{\text{Tx2,} v \comma c'} \mathbf{G}_{v \comma c}^* \mathbf{s}_{v \comma c}^*$.

The signals from spatially and frequency multiplexed UEs propagate through wireless channels and are finally received by the BS equipped with $N$ antennas. When \emph{I/Q imbalance occurs also in the parallel RX branches of the BS}, the received signal snapshot vector $\mathbf{r}_{\text{TxRxi,} c} \in{\mathbb{C}}^{N\times1}$ at subcarrier $c$, under joint TX+RX I/Q imbalances can be expressed as
\begin{align}
	\begin{split}
	\mathbf{r}_{\text{TxRxi,} c} 
	&= \mathbf{K}_{\text{Rx1,} c} \mathbf{r}_{\text{Txi,} c} +\mathbf{K}_{\text{Rx2,} c} \mathbf{r}_{\text{Txi,} c'}^* 
	\\
	&= \sum_{u=1}^{U} \widetilde{\mathbf{\Psi}}_{u \comma c} \mathbf{G}_{u \comma c} \mathbf{x}_{u \comma c}
	+ \sum_{v=1}^{V} \widetilde{\mathbf{\Omega}}_{v \comma c} \mathbf{G}_{v \comma c'}^* \mathbf{x}_{v \comma c'}^*
	\\
	&\quad+ \mathbf{K}_{\text{Rx1,} c} \mathbf{z}_c + \mathbf{K}_{\text{Rx2,} c} {\mathbf{z}}_{c'}^*
	\end{split}
	\label{r_TxRxi}
\end{align}%
where perfect time and frequency synchronization between the UEs and BS is assumed for simplicity. Here, $\widetilde{\mathbf{\Psi}}_{u \comma c} \in {\mathbb{C}}^{N \times M_u}$ and $\widetilde{\mathbf{\Omega}}_{v \comma c} \in {\mathbb{C}}^{N \times M_v}$ denote the total effective linear channels of UE $u$ and $v$ including the joint effects of TX and RX I/Q imbalances as well as the wireless propagation channels. The matrices are given by
\begin{align}
	\begin{split}
	\widetilde{\mathbf{\Psi}}_{u \text{,} c} &=
		\begin{bmatrix}
			\mathbf{K}_{\text{Rx1,} c} 
			&\mathbf{K}_{\text{Rx2,} c}
		\end{bmatrix} 
		\begin{bmatrix}
			\mathbf{H}_{u \text{,} c} &\mathbf{0} \\
			\mathbf{0}	&\mathbf{H}_{u \text{,} c'}^* 
		\end{bmatrix}
		\begin{bmatrix}
			\mathbf{K}_{\text{Tx1,} u \text{,} c} \\
			\mathbf{K}_{\text{Tx2,} u \text{,} c'}^*
		\end{bmatrix},
	\\[6pt]
	\widetilde{\mathbf{\Omega}}_{v \text{,} c} &= 
		\begin{bmatrix}
			\mathbf{K}_{\text{Rx1,} c} 
			&\mathbf{K}_{\text{Rx2,} c} 
		\end{bmatrix} 
		\begin{bmatrix}
			\mathbf{H}_{v \text{,} c} &\mathbf{0} \\
			\mathbf{0}	&\mathbf{H}_{v \comma c'}^* 
		\end{bmatrix}
		\begin{bmatrix}
			\mathbf{K}_{\text{Tx2,} v \comma c} \\
			\mathbf{K}_{\text{Tx1,} v \comma c'}^*
		\end{bmatrix}
	\end{split}
	\label{eq:PsiOmega}
\end{align}
where $\mathbf{H}_{u \comma c} \in {\mathbb{C}}^{N \times M_u}$ and $\mathbf{H}_{v \comma c} \in {\mathbb{C}}^{N \times M_v}$ are the channel response matrices of user $u$ and $v$ at subcarrier $c$, respectively, and again $c'$ denotes the image subcarrier. Throughout the paper, the channel response elements are assumed to be constants within each narrow subcarrier. Additionally, the external interference plus noise vector $\mathbf{z}_c \in {\mathbb{C}}^{N\times 1}$ at RX input is given by
\begin{align}
	{\mathbf{z}}_c = \sum_{l=1}^{L} {\mathbf{H}}_{\text{int,} l \comma c} \mathbf{s}_{\text{int,} l \comma c}+ {\mathbf{n}}_c
	\label{z_def}
\end{align}
where $\mathbf{H}_{\text{int,} l \comma c} \in{\mathbb{C}}^{N \times J_l}$ represents the channel response matrix of the $l^{\text{th}}$ interferer at subcarrier $c$. Since the interferers are generally not synchronized with the BS and since we are not limiting the study to any specific interference waveform, $\mathbf{s}_{\text{int,} l \comma c}$ is basically the result of the sampled interference signal at the desired subcarrier after the RX FFT processing. In practice, the interfering signals themselves can be modeled, e.g., with complex Gaussian signals but it should be noted that in any case, the contribution of each interferer has a strong spatial response through channel $\mathbf{H}_{\text{int,} l \comma c}$. The noise vector ${{\mathbf{n}}_c \in {\mathbb{C}}^{N\times 1}}$ models the additive noise in the RX electronics. Noise elements in different RX branches, without I/Q imbalances, are assumed to be complex circular and mutually uncorrelated. A corresponding formulation for ${\mathbf{z}}_{c'}$, i.e., the external interference and noise at the image subcarrier, is obtained from (\ref{z_def}) by substituting the subcarrier index $c$ with $c'$. 

\begin{figure}[t]
	\centering
	\includegraphics[width=\figurewidth] {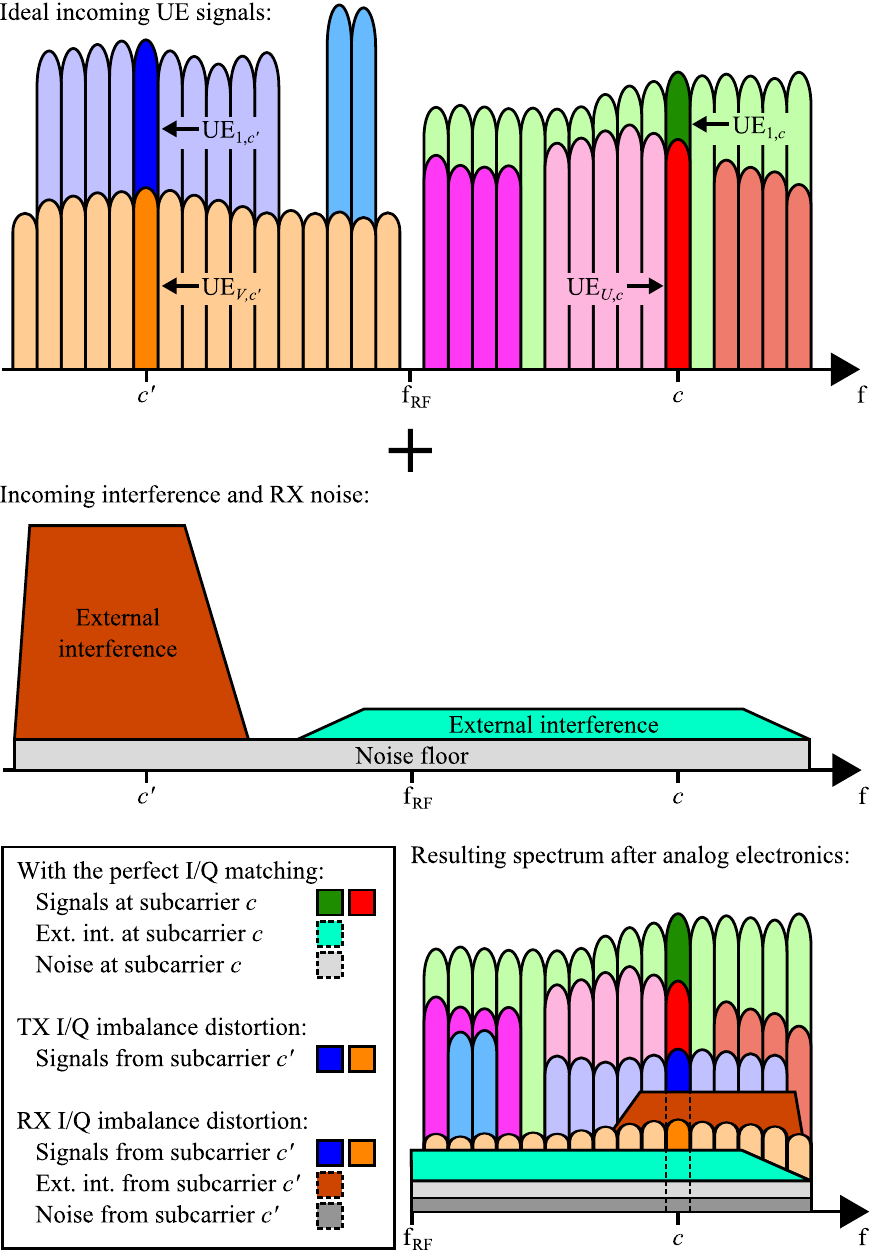}
	\caption{Illustration of the spectral components of the received signal under I/Q imbalances. The signals at the image subcarrier alias due to the both TX and RX I/Q imbalances whereas the external interference and noise are affected only by RX I/Q imbalance.}	
	\vspace{-10pt}
	\label{aliasing}
\end{figure}

The model in (\ref{r_TxRxi}) explicitly describes how the received signal is structured in OFDMA MU-MIMO systems under transceiver I/Q imbalances. Unlike in SU-MIMO systems, the received signal in (\ref{r_TxRxi}) includes substantial inter-user interference from the other spatially multiplexed UEs at subcarrier $c$. Furthermore, we note that the signal includes contribution not only from subcarrier $c$ but also from the image subcarrier $c'$. The UE signals transmitted at the image subcarrier leak to the considered subcarrier due to both TX and RX I/Q imbalances and consequently we call it inter-user interference from the image subcarrier. In contrast to UE signals, the external interference and noise alias to subcarrier $c$ only due to RX I/Q imbalance. The overall spectral structure of the received signal is illustrated in Fig. \ref{aliasing}. 

In practice, a subcarrier with very high interference levels could be left unused for data transmission if there are subcarriers with better conditions available. However, due to RX I/Q imbalance, the strong external interference from the image subcarrier aliases on top of the desired signal, even if the image subcarrier is not used for data transmission at all. This is unavoidable, since the analog electronics in transceivers are never ideal in practice. The overall scale of the signal distortion is naturally determined by the I/Q imbalance parameters, and the overall severity depends on the sensitivity of the application under consideration. Furthermore, the primary target of separating the multiplexed streams of different UEs at subcarrier $c$, under the external interference and transceiver I/Q imbalances is a key concern. This will be elaborated in the forthcoming sections where RX spatial processing is taken into account.

Note that (\ref{r_TxRxi}) expresses the received signal under joint TX+RX I/Q imbalances in a generic form. Throughout the paper, the special case with I/Q imbalance only in the TXs is obtained from the signal models by substituting $\mathbf{K}_{\text{Rx1,} c} = \mathbf{I}$ and $\mathbf{K}_{\text{Rx2,} c} = \mathbf{0}$ for all $c$. Similarly, the case with I/Q imbalance only in the RX is obtained by substituting $\mathbf{K}_{\text{Tx1,} i \comma j} = \mathbf{I}$ and $\mathbf{K}_{\text{Tx2,} i \comma j} = \mathbf{0}$ for all $i \in \{u,v\}$ and $j \in \{c,c'\}$.

\subsection{Spatial Post-Processing with Digital Combiners}

Multiple RX antennas enable flexible combining of the antenna signals for obtaining the desired system performance. Usually, the combining process is implemented by digital signal processing due to its high computational power, reconfigurability and small physical size. Generally speaking, a digital linear combiner processes the received signal snapshots with complex weights ${{\mathbf{w}} = [w_1, w_2, ... , w_N]\TT} \in {\mathbb{C}}^{N\times1}$ yielding an output signal $y= {\mathbf{w}}\HH{\mathbf{r}}$~\cite{litva_digital_1996}. When applying this method to MU-MIMO systems utilizing OFDMA waveforms, each of $Q_{u \comma c}$ transmitted data streams of each $U$ UEs at subcarrier $c$ needs an individual weight vector $\mathbf{w}_{q \comma u \comma c}$ for separating data streams from each others in the RX. In general, the weights can be selected with blind or non-blind methods, depending on \textit{a priori} information, under a given optimization criteria. The basic approach is, however, to combine the received signals from different RX branches coherently while trying to minimize the effect of the non-desired interference and noise. Since this classical processing is done at the subcarrier level, we call it per-subcarrier combiner. 

When stacking the weight vectors of individual data streams into a matrix, we get the complete weight matrix $\mathbf{W}_c~=~[\mathbf{w}_{1 \comma 1 \comma c}, \cdots, \mathbf{w}_{Q_U \comma U \comma c}] \in {\mathbb{C}}^{N \times S}$ where $S = \sum_{u=1}^{\text{U}} Q_{u \comma c}$ is the total number of the transmitted data streams at subcarrier $c$. Under joint TX+RX I/Q imbalances, based on (\ref{r_TxRxi}), the output signal vector $\mathbf{y}_{\text{TxRxi,} c} \in {\mathbb{C}}^{S \times 1}$ reads
\begin{align}
	\mathbf{y}_{\text{TxRxi,} c}&= {\mathbf{W}}_{c}\HH {\mathbf{r}_{\text{TxRxi,} c}} 
	\nonumber \\
	&= \sum_{u=1}^{U} {\mathbf{W}}_{c}\HH \widetilde{\mathbf{\Psi}}_{u \comma c} \mathbf{G}_{u \comma c} \mathbf{x}_{u \comma c}
	+ \sum_{v=1}^{V} {\mathbf{W}}_{c}\HH \widetilde{\mathbf{\Omega}}_{v \comma c} \mathbf{G}_{v \comma c'}^* \mathbf{x}_{v \comma c'}^*
	\nonumber \\
	&\quad+ {\mathbf{W}}_{c}\HH \mathbf{K}_{\text{Rx1,} c} {\mathbf{z}}_c 
	+ {\mathbf{W}}_{c}\HH \mathbf{K}_{\text{Rx2,} c} {\mathbf{z}}_{c'}^*.
\label{y_TxRxi}
\end{align} 
The entries of the output signal vector $\mathbf{y}_{\text{TxRxi,} c}$ represent the data streams originating from different UEs and are thus forwarded to further signal processing stages such as decoding. As visible in~(\ref{y_TxRxi}), all signal terms are multiplied with the same weighting matrix. The first term contains the data streams of all $U$ desired UEs, which are to be separated by the spatial processing, while at the same time suppressing the effects of the other terms as much as possible. The second term is due to the inter-user interference from the mirror UEs in the OFDMA framework while the third and fourth terms are due to the external interference and noise. Notice that the external interference contributes to the combiner output through direct co-channel coexistence as well as due to image subcarrier leakage. The above is clearly a challenge when optimizing the combiner weights and it becomes even more difficult when the number of multiplexed UEs and external interferers is increased. 

\begin{figure}[t]
	\centering
	\includegraphics[width=\figurewidth]{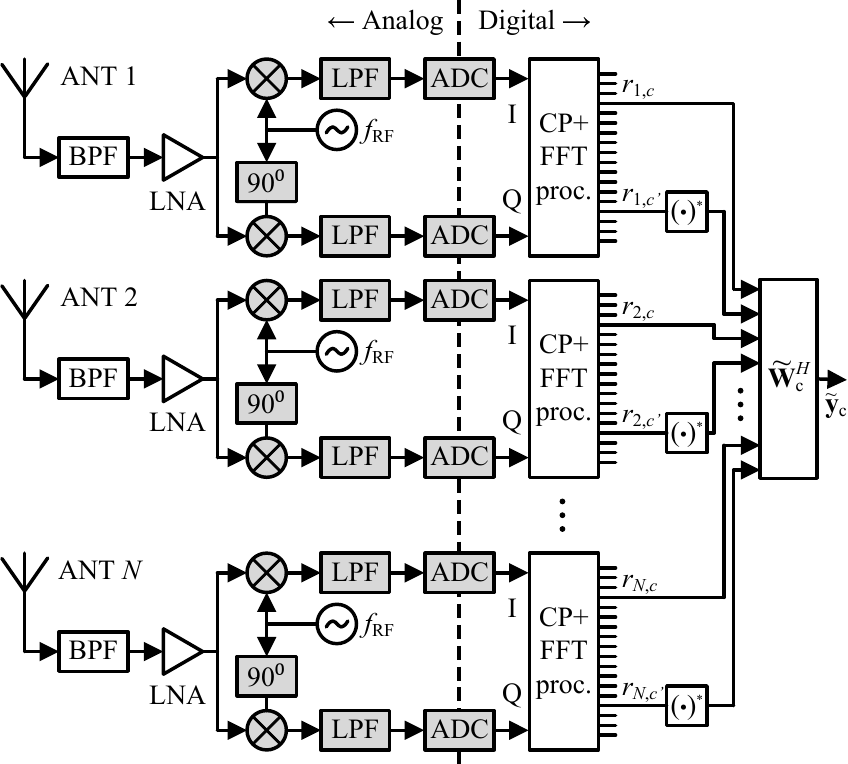}
	\vspace{6pt}
	\caption{The augmented RX combining in MU-MIMO systems utilizing OFDMA waveforms. The blocks with gray shading represent the main sources of I/Q imbalance in RX.}
	\label{combining_concept}
\end{figure}

Since transceiver I/Q imbalances cause both inter-user interference and external interference through image subcarrier leakage, classical per-subcarrier spatial processing can easily run out of degrees of freedom to suppress all of them sufficiently. To alleviate this and enhance the interference suppression capabilities, we next augment the spatial combiner operating principle to process each subcarrier along with its image subcarrier jointly. This means augmented processing where the signals from both subcarriers $c$ and $c'$ are combined with two separate sets of weights as illustrated in Fig. \ref{combining_concept}. The approach has been shown to be efficient for I/Q imbalance mitigation in SU-MIMO communication with OFDM waveforms~\cite{schenk_estimation_2006, tarighat_mimo_2005, ozdemir_i/q_2013}. In this paper, we extend the augmented combiner to operate in the considerably more challenging MU-MIMO OFDMA scheme including also the influence of the external interference.

We denote the weight sets of the augmented combiner for transmitted data stream $q$ of user $u$ at subcarriers $c$ and $c'$ by $\mathbf{w}_{\text{A,} q \comma u \comma c} \in {\mathbb{C}}^{N \times 1}$ and \mbox{$\mathbf{w}_{\text{B,} q \comma u \comma c'} \in {\mathbb{C}}^{N \times 1}$}, and stack them into the augmented weight vector \mbox{$\widetilde{\mathbf{w}}_{q \comma u \comma c} = [\mathbf{w}_{\text{A,} q \comma u \comma c}\TT, \mathbf{w}_{\text{B,} q \comma u \comma c'}\TT]\TT \in {\mathbb{C}}^{2N\times 1}$}. Then, similarly as for the per-subcarrier processing, the weights of individual data streams are stacked, resulting in the complete augmented weight matrix given by \mbox{$\widetilde{\mathbf{W}}_c = [\widetilde{\mathbf{w}}_{1 \comma 1 \comma c}, \cdots, \widetilde{\mathbf{w}}_{Q_U \comma U \comma c}] \in {\mathbb{C}}^{2N \times S}$. After} defining the augmented signal vector under joint TX+RX I/Q imbalances as $\widetilde{\mathbf{r}}_{\text{TxRxi,} c} = [\mathbf{r}_{\text{TxRxi,} c}\TT, \mathbf{r}_{\text{TxRxi,} c'}\HH ]\TT \in {\mathbb{C}}^{2N \times 1}$ where $\mathbf{r}_{\text{TxRxi,} c}$ is as given in (\ref{r_TxRxi}), the output signal of the augmented combiner under joint TX+RX I/Q imbalances becomes
\begin{align}
	\widetilde{\mathbf{y}}_{\text{TxRxi,} c} &= \widetilde{\mathbf{W}}_c\HH \widetilde{\mathbf{r}}_{\text{TxRxi,} c} 
	\nonumber \\
	&= \sum_{u=1}^{U} \widetilde{\mathbf{W}}_c\HH \widetilde{\mathbf{\Xi}}_{u \comma c} \mathbf{G}_{u \comma c} \mathbf{x}_{u \comma c}
	+ \sum_{v=1}^{V} \widetilde{\mathbf{W}}_c\HH \widetilde{\mathbf{\Phi}}_{v \comma c} \mathbf{G}_{v \comma c'}^* \mathbf{x}_{v \comma c'}^*
	\nonumber \\
	&\quad + \widetilde{\mathbf{W}}_c\HH \widetilde{\mathbf{K}}_{\text{RxA,} c} {\mathbf{z}}_c 
	+ \widetilde{\mathbf{W}}_c\HH \widetilde{\mathbf{K}}_{\text{RxB,} c} {\mathbf{z}}_{c'}^*
	\label{y_WL_TxRxi}
\end{align} 
where the effective augmented channel matrices $\widetilde{\mathbf{\Xi}}_{u \comma c} \in {\mathbb{C}}^{2N \times M_u}$ and $\widetilde{\mathbf{\Phi}}_{v \comma c} \in {\mathbb{C}}^{2N \times M_v}$ are equal to
\begin{align}
	\begin{split}
	\widetilde{\mathbf{\Xi}}_{u \comma c} = 
		\begin{bmatrix}
			\mathbf{K}_{\text{Rx1,} c} &\mathbf{K}_{\text{Rx2,} c} \\
			\mathbf{K}_{\text{Rx2,} c'}^* &\mathbf{K}_{\text{Rx1,} c'}^*
		\end{bmatrix} 
		\begin{bmatrix}
			\mathbf{H}_{u \comma c} &\mathbf{0} \\
			\mathbf{0}	&\mathbf{H}_{u \comma c'}^* 
		\end{bmatrix}
		\begin{bmatrix}
			\mathbf{K}_{\text{Tx1,} u \comma c} \\
			\mathbf{K}_{\text{Tx2,} u \comma c'}^*
		\end{bmatrix},
	\\[6pt]
	\widetilde{\mathbf{\Phi}}_{v \comma c} = 
		\begin{bmatrix}
			\mathbf{K}_{\text{Rx1,} c} &\mathbf{K}_{\text{Rx2,} c} \\
			\mathbf{K}_{\text{Rx2,} c'}^* &\mathbf{K}_{\text{Rx1,} c'}^*
		\end{bmatrix} 
		\begin{bmatrix}
			\mathbf{H}_{v \comma c} &\mathbf{0} \\
			\mathbf{0}	&\mathbf{H}_{v \comma c'}^* 
		\end{bmatrix}
		\begin{bmatrix}
			\mathbf{K}_{\text{Tx2,} v \comma c} \\
			\mathbf{K}_{\text{Tx1,} v \comma c'}^*
		\end{bmatrix}.
	\end{split}
\end{align}
Moreover, the augmented RX I/Q imbalance matrices $\widetilde{\mathbf{K}}_{\text{RxA,} c}$ and $\widetilde{\mathbf{K}}_{\text{RxA,} c}$, both $\in {\mathbb{C}}^{2N \times N}$, are equal to
\begin{align}
	\widetilde{\mathbf{K}}_{\text{RxA,} c} = 
	\begin{bmatrix}
		\mathbf{K}_{\text{Rx1,} c} \\
		\mathbf{K}_{\text{Rx2,} c'}^*
	\end{bmatrix} \comma \quad
	\widetilde{\mathbf{K}}_{\text{RxB,} c} = 
	\begin{bmatrix}
		\mathbf{K}_{\text{Rx2,} c} \\
		\mathbf{K}_{\text{Rx1,} c'}^*
	\end{bmatrix}.
	\label{K_Rx_A_B}
\end{align}
Clearly, the output signal structures of the conventional and augmented combiners are very similar. However, the underlying difference is that (\ref{y_WL_TxRxi}) adopts twice as many weights as (\ref{y_TxRxi}) for processing signals at subcarriers $c$ and $c'$ jointly. Naturally, this doubles the computational complexity of the combining process but also gives us more degrees of freedom for obtaining the desired signal separation and interference suppression, even under challenging I/Q imbalances. Note that this flexibility is achieved by modifying the combiner block only whereas the costly RF chains and demanding FFT processing remain the same as in per-subcarrier processing. Notice also that various kinds of special cases, e.g., TX I/Q imbalances only or RX I/Q imbalances only, are naturally obtained as corresponding special cases of (\ref{y_WL_TxRxi})--(\ref{K_Rx_A_B}) by proper substitutions.

\section{LMMSE and Augmented LMMSE Receivers and Output SINRs}

In this section, we derive the MMSE optimal linear and augmented combiners under TX+RX I/Q imbalances. In addition, we seek to characterize the output performance of the spatial combiners in terms of the combiner output SINR. Finally, we show how much the computational complexity of the whole digital signal processing chain is increased when using the augmented method instead of the more ordinary linear counterpart.

\subsection{Received Signal Covariance Matrix}

First, to support the upcoming RX derivations and SINR expressions, we derive an expression for the covariance matrix of the augmented received signal vector under joint TX+RX I/Q imbalances. We assume that the data streams of different UEs, the data streams at subcarriers $c$ and $c'$ (with perfect I/Q matchings), the interfering signals and the additive noise are all mutually uncorrelated. In addition, we assume that the external interference as well as noise at RX input are complex circular. Finally, we assume that all TX data streams of an individual UE $u$ have equal powers before the TX stream-to-antenna mapping with precoder $\mathbf{G}_{u \comma c}$.

Under the assumptions above, the covariance matrix $\widetilde{\mathbf{R}}_{\text{TxRxi,} c} \in {\mathbb{C}}^{2N\times 2N}$ of the \emph{augmented signal model} under joint TX+RX I/Q imbalances can be expressed as
\begin{align}
	\begin{split}
	\widetilde{\mathbf{R}}_{\text{TxRxi,} c}
	&= \E \left[ \widetilde{\mathbf{r}}_{\text{TxRxi,} c} \widetilde{\mathbf{r}}_{\text{TxRxi,} c}\HH \right] \\
	&= \sum_{u=1}^U \sigma_{\text{x,} u \comma c}^2 \widetilde{\mathbf{\Xi}}_{u \comma c} \mathbf{G}_{u \comma c} \mathbf{G}_{u \comma c}\HH \widetilde{\mathbf{\Xi}}_{u \comma c}\HH 
	\\
	&\quad + \sum_{v=1}^V \sigma_{\text{x,} v \comma c'}^2 \widetilde{\mathbf{\Phi}}_{v \comma c} \mathbf{G}_{v \comma c'}^* \mathbf{G}_{v \comma c'}\TT \widetilde{\mathbf{\Phi}}_{v \comma c}\HH 
	\\
	&\quad + \widetilde{\mathbf{K}}_{\text{RxA,} c} {\mathbf{R}}_{\text{z,} c} \widetilde{\mathbf{K}}_{\text{RxA,} c}\HH
	+ \widetilde{\mathbf{K}}_{\text{RxB,} c} {\mathbf{R}}_{\text{z,} c'}^* \widetilde{\mathbf{K}}_{\text{RxB,} c}\HH
	\end{split}
	\label{R_WL_TxRxi}
\end{align}%
where $\sigma_{\text{x,} u \comma c}^2 = \E [ | x_{u \comma  c} |^2 ]$ denotes the power of an individual data stream of user $u$ at subcarrier $c$. In addition, the covariance matrix of the external interference plus noise, $\mathbf{R}_{\text{z,} c} \in {\mathbb{C}}^{N\times N}$, equals
\begin{align}
	\mathbf{R}_{\text{z,} c} = \E \left[ {\mathbf{z}}_c {\mathbf{z}}_c\HH \right]
	= \sum_{l=1}^L \sigma_{\text{int,} l \comma c}^2\mathbf{H}_{\text{int,} l \comma c}\mathbf{H}_{\text{int,} l \comma c}\HH 
	+ \sigma_{\text{n,} c}^2 \mathbf{I}
	\label{R_z}
\end{align}
where $\sigma_{\text{int,} l \comma c}^2$ denotes the power of the $l^{\text{th}}$ external interferer and $\sigma_{\text{n,} c}^2$ denotes the noise power, both at subcarrier $c$. In general, the covariance matrix of the received signal has a very intuitive structure since it depends directly on the stream powers, channel matrices, and the external interference and noise. This kind of covariance structure is, in principle, well-known in the literature. However, I/Q imbalances cause signal leakage from the image subcarrier and thus generate additional terms to the covariance matrix, i.e., the second and fourth terms in (\ref{R_WL_TxRxi}). In addition, the propagation responses are modified from pure wireless channels to total effective channels including also the effects of the TX and RX electronics. The practical consequences of this kind of distortion will be quantified next in Section III.B. Notice that as a special case, the covariance matrix for the linear (non-augmented) signal model in (\ref{r_TxRxi}) is given as the first quadrant of (\ref{R_WL_TxRxi}).

\subsection{Signal-to-Interference-plus-Noise Ratio (SINR)}

Next, we quantify the performance of the augmented combiner output signal under I/Q imbalances in terms of {\it the instantaneous SINR for an arbitrary data stream $q$ originating from an arbitrary UE $u$}. In general, the total output power of an arbitrary data stream $q$ of UE $u$ is equal to
\begin{align}
	\widetilde{P}_{q \comma u \comma c} 
	= \E \left[ \left| \widetilde{y}_{\text{TxRxi,} q \comma u \comma c} \right|^2 \right] 
	= \widetilde{\mathbf{w}}_{q \comma u \comma c}\HH \widetilde{\mathbf{R}}_{\text{TxRxi,} c} \widetilde{\mathbf{w}}_{q \comma u \comma c} 
	\label{P_WL_q}
\end{align}
where, $\widetilde{\mathbf{w}}_{q \comma u \comma c}$ refers to the augmented combiner weight vector corresponding to data stream $q$ of UE $u$ at subcarrier $c$ and is easily obtained by selecting the corresponding column from the weight matrix $\widetilde{\mathbf{W}}_c$. In order to express the SINR for this arbitrary data stream, we split the combiner output power in (\ref{P_WL_q}) to useful signal and interference/noise terms as follows:
\begin{align}
	\widetilde{P}_{q \comma u \comma c} &= 
	\widetilde{P}_{x \comma q \comma u \comma c} + \widetilde{P}_{\text{ISI} \comma q \comma u \comma c} 
	+ \widetilde{P}_{\text{IUI} \comma u \comma c} + \widetilde{P}_{\text{IUI} \comma c'} 
	+ \widetilde{P}_{\text{z,} c} + \widetilde{P}_{\text{z,} c'}.
	\label{P_WL_separated}
\end{align}
Here, $\widetilde{P}_{x \comma q \comma u \comma c}$ denotes the output power of the desired data stream $q$ and $\widetilde{P}_{\text{ISI} \comma q \comma u \comma c}$ represents the effect of the inter-stream interference originating from the other streams of the same UE $u$. These terms are both originating from UE $u$, but they are separated because when examining the received signal from an individual but arbitrary stream $q$ of UE $u$ perspective, the other streams of the same UE are also treated as interference. In addition, $\widetilde{P}_{\text{IUI} \comma u \comma c}$ and $\widetilde{P}_{\text{IUI} \comma c'}$ represent the inter-user interference from subcarriers $c$ and $c'$, respectively. Finally, $\widetilde{P}_{\text{z} \comma c}$ and $\widetilde{P}_{\text{z} \comma c'}$ denote the output powers of the external interference and noise originating from subcarriers $c$ and $c'$, respectively. The detailed derivations for these power terms are given in \nameref{app1}. Then for the augmented signal model, {\it the instantaneous SINR of a single received data stream under joint TX+RX I/Q imbalances} can be expressed straightforwardly by
\begin{align}
	\widetilde{\text{SINR}}_{\text{TxRxi,} q \comma u \comma c} 
	= \frac{ \widetilde{P}_{x \comma q \comma u \comma c}}
	{ \vphantom{\widetilde{\widetilde{P}_{\text{ISI} \comma c}}} \widetilde{P}_{\text{ISI} \comma q \comma u \comma c}
	+ \widetilde{P}_{\text{IUI} \comma u \comma c}
	+ \widetilde{P}_{\text{IUI} \comma c'}
	+ \widetilde{P}_{\text{z,} c} 
	+ \widetilde{P}_{\text{z,} c'}} .
	\label{SINR_WL_TxRxi}
\end{align}
Note that this per-data-stream SINR includes the effects of the RX spatial processing with given, yet arbitrary, combiner weights, while the actual derivation of the linear and augmented linear MMSE optimum coefficients is provided in the next subsection. Furthermore, it should be noted that the SINR averaged over the channel fading distribution is, in general, given by $\overline{\text{SINR}}_{\text{TxRxi,} q \comma u \comma c} = \E_{\mathbf{h}} [\widetilde{\text{SINR}}_{\text{TxRxi,} q \comma u \comma c}]$ where $\E_{\mathbf{h}}[\cdot]$ denotes statistical expectation over all associated channel fading variables and $\mathbf{h}$ is composed of all non-zero elements in $\mathbf{H}_{u\comma c}$, $\mathbf{H}_{u\comma c'}$, $\mathbf{H}_{v\comma c}$, $\mathbf{H}_{v\comma c'}$, $\mathbf{H}_{\text{int,} l\comma c}$, $\mathbf{H}_{\text{int,} l\comma c'}, \forall u, v, l$. When expanding the power terms in (\ref{SINR_WL_TxRxi}), given in \nameref{app1}, it becomes evident that deriving an exact closed-form expression for the SINR averaged over the fading variables in the considered MU-MIMO scenario is infeasible due to the intractable algebraic representation, unlike in the more simple SU-SISO OFDM scheme~\cite{narasimhan_digital_2009, ozdemir_exact_2014}, in the SU-MISO OFDM scheme~\cite{ozdemir_i/q_2013}, or in MU-SISO SC-FDMA schemes~\cite{gomaa_multi-user_2011, gomaa_phase_2014} with only one active UE at each subcarrier. Thus, in Section IV, we provide comprehensive numerical results where the above SINR expression is numerically averaged across all fading variables through Monte-Carlo simulations.

In general, the inter-stream and inter-user interferences as well as the external interference, all at subcarrier $c$, are unavoidable in the RX antennas of the considered OFDMA MU-MIMO systems but their effects on the output signal can be suppressed to some extent through antenna array processing. Based on (\ref{SINR_WL_TxRxi}), I/Q imbalances in general cause substantial SINR degradation due to signal leakage from the image subcarrier, i.e., the performance is deteriorated also by the inter-user interference and external interferences from the image subcarrier. Such a phenomenon obviously limits the overall performance and sets additional requirements for the combiner weight optimization task which will be considered in the next subsection. We emphasize that the existing works in the literature, such as~\cite{tarighat_compensation_2005, tarighat_joint_2007, ozdemir_exact_2013, ozdemir_exact_2014, krone_capacity_2008, narasimhan_digital_2009, maham_impact_2012, ozdemir_sinr_2012, tarighat_mimo_2005, schenk_estimation_2006, schenk_performance_2007, schenk_rf_2008, ozdemir_i/q_2013, qi_compensation_2010, tandur_joint_2007, gomaa_multi-user_2011, yoshida_analysis_2009, gomaa_phase_2014}, have not considered the effects of the external interference and spatially multiplexed UEs causing inter-user interference.

The corresponding power and SINR expressions for the more conventional per-subcarrier combiner are obtained from (\ref{P_WL_q})--(\ref{SINR_WL_TxRxi}) and (\ref{eq:power_terms1})--(\ref{eq:power_terms6}) by substituting the augmented quantities by their linear counterparts but are not shown explicitly due to space constraints. Furthermore, the special case with I/Q imbalance only in the TXs is obtained from all expressions by substituting $\mathbf{K}_{\text{Rx1,} c} = \mathbf{I}$ and $\mathbf{K}_{\text{Rx2,} c} = \mathbf{0}$ for all $c$. Similarly, the case with I/Q imbalance only in the RX is obtained by substituting $\mathbf{K}_{\text{Tx1,} i \comma j} = \mathbf{I}$ and $\mathbf{K}_{\text{Tx2,} i \comma j} = \mathbf{0}$ for all $i \in \{u,v\}$ and $j \in \{c,c'\}$. We will illustrate and discuss the influence of different I/Q imbalance scenarios on the practical SINR performance in more detail in Section \ref{sec:simulations}.

\subsection{Linear and Augmented Linear MMSE Combiners}

The above SINR expression is in principle valid for any possible combiner coefficients, while the optimization of the combiner coefficients is addressed next. A well-known statistical method for solving stationary estimation problems is the so-called Wiener filter which yields the optimal linear solution in the MMSE sense~\cite{trees_detection_2004}. We have shown in~\cite{hakkarainen_interference_2014} that the Wiener filter approach, when generalized to augmented or widely-linear processing, can be successfully used for the channel and hardware characteristic estimation problem under I/Q imbalance in SU-SIMO systems. Here this simple and intuitive approach is extended to cover the weight selection problem in the considered MU-MIMO OFDMA systems whereas other weight optimization methods could be used as well.

We first define the ordinary Wiener filter or LMMSE weights for signal model (\ref{r_TxRxi}), i.e., under joint TX+RX I/Q imbalances. The weights are of the form 
\begin{align}
	\mathbf{W}_{\text{TxRxi,} c}^{\text{LMMSE}} = \mathbf{R}_{\text{TxRxi,} c}^{-1} \mathbf{V}_{\text{TxRxi,} c}
	\label{wiener_ideal}
\end{align}%
where $\mathbf{R}_{\text{TxRxi,} c} = \E [ \mathbf{r}_{\text{TxRxi,} c} \mathbf{r}_{\text{TxRxi,} c}\HH ] \in {\mathbb{C}}^{N \times N}$ denotes the covariance matrix of the received signal and is equal to top-left quadrant of $\widetilde{\mathbf{R}}_{\text{TxRxi,} c}$ in (\ref{R_WL_TxRxi}). In addition, $\mathbf{V}_{\text{TxRxi,} c} = [\mathbf{v}_{\text{TxRxi,} 1 \comma 1 \comma c}, \cdots, \mathbf{v}_{\text{TxRxi,} Q_U \comma U \comma c}] \in {\mathbb{C}}^{N \times S}$ is a matrix consisting of the cross-correlation vectors between the received signal snapshots and transmitted data streams. Under joint TX+RX I/Q imbalances the cross-correlation vector, related to data stream $q$ of user $u$, is easily shown to read
\begin{align}
	\mathbf{v}_{\text{TxRxi,} q \comma u \comma c} &= \E \left[ \mathbf{r}_{\text{TxRxi,} c} x_{q \comma u \comma c}^* \right] 
	= \sigma_{x \comma u \comma c}^2 \widetilde{\mathbf{\Psi}}_{u \comma c} \mathbf{G}_{u \comma c} \mathbf{e}_{q}.
	\label{v_TxRxi}
\end{align}%
Now, the combiner weight vector related to data stream $q$ of UE $u$ at subcarrier $c$ is equal to 
\begin{align}
	\mathbf{w}_{\text{TxRxi,} q \comma u \comma c}^{\text{LMMSE}} = \mathbf{R}_{\text{TxRxi}, c}^{-1} \mathbf{v}_{\text{TxRxi,} q \comma u \comma c}.
	\label{row_weight_imb}
\end{align}%
Notice that if I/Q imbalances are set to zero, (\ref{row_weight_imb}) reduces to the classical Wiener filter as expected.

We next proceed to the augmented combiner coefficient optimization in the MMSE sense, referred to as augmented LMMSE or augmented Wiener filter in the following. The weight optimization problem under joint TX+RX imbalances corresponds to solving the augmented weights as
\begin{align}
	\widetilde{\mathbf{W}}_{\text{TxRxi,} c}^{\text{LMMSE}} = \widetilde{\mathbf{R}}_{\text{TxRxi,} c}^{-1} \widetilde{\mathbf{V}}_{\text{TxRxi,} c}
\end{align}
where $\widetilde{\mathbf{R}}_{\text{TxRxi,} c}$ is given in (\ref{R_WL_TxRxi}) and $\widetilde{\mathbf{V}}_{\text{TxRxi,} c} = [\widetilde{\mathbf{v}}_{\text{TxRxi,} 1 \comma 1 \comma c}, \cdots, \widetilde{\mathbf{v}}_{\text{TxRxi,} Q_U \comma U \comma c}] \in {\mathbb{C}}^{2N \times S}$ is the cross-correlation matrix. There, the column vector related to data stream $q$ of UE $u$ at subcarrier $c$ reads
\begin{align}
	\begin{split}
	\widetilde{\mathbf{v}}_{\text{TxRxi,} q \comma u \comma c} &= \E \left[ \widetilde{\mathbf{r}}_{\text{TxRxi,} c} x_{q \comma u \comma c}^*  \right] \\
	&= \begin{bmatrix}
		\sigma_{x \comma u \comma c}^2 \widetilde{\mathbf{\Psi}}_{u \comma c} \mathbf{G}_{u \comma c} \mathbf{e}_{q}
		\\[6pt]
		\sigma_{x \comma u \comma c}^2 \widetilde{\mathbf{\Omega}}_{u \comma c'}^* \mathbf{G}_{u \comma c} \mathbf{e}_{q}		
	\end{bmatrix} \\
	&= \sigma_{x \comma u \comma c}^2 \widetilde{\mathbf{\Xi}}_{u \comma c} \mathbf{G}_{u \comma c} \mathbf{e}_{q}
	\label{v_WL_TxRxi}
	\end{split}
\end{align}%
and the augmented weight vector related to the same data stream becomes consequently equal to
\begin{align}
	\widetilde{\mathbf{w}}_{q \comma u \comma c}^{\text{LMMSE}} = \widetilde{\mathbf{R}}_{\text{TxRxi}, c}^{-1} \widetilde{\mathbf{v}}_{\text{TxRxi,} q \comma u \comma c}.
	\label{row_weight_WL_imb}
\end{align}%
This kind of processing methods, which can efficiently suppress not only the classical mirror-subcarrier crosstalk within a single UE but more generally the inter-user interference inside a subcarrier and between mirror-subcarriers, as well as the external interference and noise, in spite of substantial I/Q imbalances in multi-antenna TXs and RXs, will play a big role in MU-MIMO networks especially in interference-limited conditions. 
Moreover, the derived augmented combiner can provide good performance also under reasonable levels of carrier frequency offsets (CFOs) and timing offsets, under the following assumptions. In practical systems the maximum CFO between the UEs and the BS is typically relatively small. E.g., it is said in the LTE/LTE-Advanced specifications \cite{3gpp_evolved_2014, 3gpp_evolved_2013} that {\it "the UE modulated carrier frequency shall be accurate within} $\pm0.1$~PPM {\it observed over a period of one time slot} (0.5~ms) {\it compared to the carrier frequency received from the E-UTRA Node B.''} With 2~GHz carrier frequency, as a concrete example, this would mean a CFO equal to maximum of $\pm$~200~Hz which is very small relative to 15~kHz subcarrier spacing (maximum of 1.3\%). Therefore, the resulting inter-carrier interference and inter-user interference are also very small. Also, the simplest approximation of the CFO effects is that small CFOs map only to common phase error per UE and per OFDMA symbol being thus a part of the effective wireless channel, and is properly handled and processed as long as the system supports regular reference signals for channel estimation. If further assuming that the timing offsets are within the cyclic prefix length for all spatially multiplexed UEs, also these offsets simply map to being part of the effective frequency selective fading channels which are structurally processed in the proposed augmented combiner. We also emphasize that there are no explicit constraints related to synchronization of the co-channel interference signals, as they are treated in the modeling and analysis as arbitrary waveforms. 

As will be illustrated by the numerical results in Section IV, the formulated augmented Wiener combiner has the structural capability to achieve substantially better performance compared to more classical per-subcarrier processing, assuming that the needed correlations devised above are available. In practice, the Wiener solutions can be well approximated by various adaptive estimation methods based on known training or reference signals~\cite{widrow_adaptive_1967}. In the next subsection, we shortly address how the selection of the weight estimation algorithm influences the overall computational complexity and achievable performance in the considered MU-MIMO scenario.

\subsection{Spatial Processing Computational Complexity Aspects}

In this subsection we focus on computing complexity aspects of the developed RX processing scheme with joint channel equalization, data stream separation and I/Q imbalance mitigation. In particular, we present the computational complexities of the three main digital signal processing blocks in the MIMO BS, namely the FFT processing, weight estimation and digital combining. All the computational complexities are here given in {\it real-valued} arithmetic operations (additions and multiplications) called floating point operations (flops) unless otherwise stated.

\subsubsection{FFT processing}

In OFDMA RXs FFT processing takes down-converted and digitized received signals as an input and converts them to subcarrier level observations. The computational complexity of $C$-point FFT with the current state-of-the-art split-radix implementation is given by~\cite{johnson_modified_2007}
\begin{align}
	\begin{split}
	\Theta_{\text{FFT}} &= \frac{34}{9}C \lgg C - \frac{124}{27} C - 2 \lgg C \\ 
	&\quad - \frac{2}{9}\left( -1 \right)^{\lgg C} \lgg C + \frac{16}{27} \left( -1 \right)^{\lgg C} + 8.
	\end{split}
	\label{comp_FFT}
\end{align}%
This is the overall complexity of the whole FFT block in a single RX branch and thus the result in (\ref{comp_FFT}) can be divided by $C$ if one seeks to quantify the computational load from a single subcarrier perspective.

\subsubsection{Weight estimation}

After the FFT processing, the receiving BS must estimate the combining weights. While the previous subsection presented the LMMSE and augmented LMMSE optimum combiner solutions, they can in practice be well estimated or approximated using reference signals together with adaptive filtering algorithms~\cite{widrow_adaptive_1967, haykin_adaptive_2002}. Here, we shortly address the computational complexities of two elementary adaptive algorithms, namely the least mean squares (LMS) and recursive least squares (RLS), while various alternative algorithms can also be adopted in practice. In order to easily compare the computational burden between the linear per-subcarrier and the augmented subcarrier processing methods, we define $N_{\text{in}}$ to denote the number of parallel input samples of the estimation algorithm. For the linear per-subcarrier processing $N_{\text{in}} = N$ whereas for the augmented combiner $N_{\text{in}} = 2N$ due to joint processing of signals from subcarriers $c$ and $c'$ as visible in Fig. \ref{combining_concept}.

The well-known form of the LMS algorithm, described in detail in~\cite[p.~238]{haykin_adaptive_2002}, requires \mbox{$2N_{\text{in}}+1$} complex-valued multiplications and $2N_{\text{in}}$ complex-valued additions per iteration round. Thus, the resulting per-data-stream complexity of LMS at a single subcarrier is given by 
\begin{align}
	\Theta_{\text{LMS,} c} = 16N_{\text{in}} + 6.
\end{align}
Correspondingly, a straightforward RLS implementation, equal to the one in~\cite[p.~442]{haykin_adaptive_2002}, requires $4N_{\text{in}}^2 + 3N_{\text{in}}$ complex-valued multiplications, $3N_{\text{in}}^2 + N$ complex-valued additions and $2N_{\text{in}}^2$ real-valued multiplications per iteration round. Consequently, the per-data-stream complexity of RLS at a single subcarrier is equal to
\begin{align}
	\Theta_{\text{RLS,} c} = 32 N_{\text{in}}^2 + 20 N_{\text{in}}.
\end{align}%
The computational complexity of RLS is clearly higher than that of LMS. Furthermore, with large $N$ the ratio of the complexity between the augmented processing and the linear per-subcarrier processing is two for LMS and four for RLS.

\subsubsection{Digital combining}

When the weights have been estimated, the BS adjusts the amplitudes and phases of the signals in different antennas branches and finally adds up all the antenna signals. These operations cause $N_{\text{in}}$ complex-valued multiplications and $N_{\text{in}} - 1$ complex-valued additions which results in the total per-data-stream computational complexity given by 
\begin{align}
	\Theta_{\text{combining,} c} &= 8 N_{\text{in}} - 2.
\end{align}
Note that also here the complexity of the augmented processing is practically doubled when compared to that of the ordinary per-subcarrier processing. 

\subsubsection{Overall Complexity and Discussion}

The signal path from the ADC outputs to the combiner output includes the FFTs, weight estimation and digital combining. Whereas the FFT processing is carried out only once for a given signal block in every RX branch, the weight estimation and combining are done separately for each data stream but jointly for all RX branches. The ratios between the overall complexities of the augmented processing and the linear per-subcarrier processing are presented in Table~\ref{tab:LMS_ratio} when the weights of an arbitrary stream are estimated with the LMS principle, and similarly in Table~\ref{tab:RLS_ratio} with the RLS approach. The results show that with LMS, the augmented processing requires 33\%--98\% more computational effort than the per-subcarrier processing. The corresponding results of RLS show an increase equal to 121\%--298\%. Interestingly, these results clearly indicate that the augmented processing is not necessarily doubling the overall complexity, as often misleadingly stated, but the increase is highly dependent on the considered scenario. In particular, the less RX antennas are involved and the less data streams need to be separated, the closer is the complexity of the augmented processing to that of the per-subcarrier processing. Moreover, when increasing the number of subcarriers, the influence of the computationally heavy FFT processing is emphasized and thus the difference between the augmented and linear processing methods decreases. 

The results and discussion above are based on the conventional implementations of the adaptive algorithms. More advanced versions of the algorithms, such as the normalized LMS~\cite[p.~324]{haykin_adaptive_2002}, would naturally change the exact results. However, our intention here is to highlight that the increase in the computational complexity of the augmented subcarrier processing can vary from only a few tens of percents to several hundreds of percents when compared to the conventional linear processing. Consequently, the selection between the augmented and conventional processing needs always careful consideration and is eventually a trade-off between the complexity and performance.

\newcolumntype{C}[1]{>{\centering\let\newline\\\arraybackslash\hspace{0pt}}b{#1}}
\begin{table}[t]
	\centering
	\caption{The ratio of the computational complexities between the augmented subcarrier processing and the linear per-subcarrier processing as a function of the number of RX antennas $N$ and the number of data streams $S$. Comparison includes FFT processing, weight estimation with the given algorithm as well as digital combining.}
	\subcaption{Weights solved by the augmented/linear LMS.}
	\label{tab:LMS_ratio}
	\begin{tabular}{ C{0.65cm} | C{0.65cm} | *{5}{| C{0.65cm}}}
		\hline \tstruta
		\multirow{2}{*}N &\multirow{2}{*}S &\multicolumn{5}{c}{FFT size} \\
     			& &64 &256 &1024 &2048 &8192\\ 
   	   		\hline \hline \tstrutc
					1 &1 &1.52 &1.45 &1.39 &1.37 &1.33\\
      		10 &5 &1.86 &1.81 &1.77 &1.75 &1.72\\
      		20 &10 &1.92 &1.90 &1.87 &1.86 &1.84\\
      		100 &50 &1.98 &1.98 &1.97 &1.97 &1.96\\
					\hline
	\end{tabular}		
	\vspace{10pt}
	\subcaption{Weights solved by the augmented/linear RLS.}
	\label{tab:RLS_ratio}
	\begin{tabular}{ C{0.65cm} | C{0.65cm} | *{5}{| C{0.65cm}}}
		\hline \tstruta
		\multirow{2}{*}N &\multirow{2}{*}S &\multicolumn{5}{c}{FFT size} \\
    			& &64 &256 &1024 &2048 &8192\\ 
      		\hline \hline \tstrutc
					1 &1 &2.63 &2.48 &2.36 &2.31 &2.21\\
      		10 &5 &3.81 &3.80 &3.79 &3.78 &3.77\\
      		20 &10 &3.91 &3.91 &3.90 &3.90 &3.90\\
      		100 &50 &3.98 &3.98 &3.98 &3.98 &3.98\\
					\hline
	\end{tabular}
\end{table}%

\section{Numerical Evaluations, Illustrations and Analysis}
\label{sec:simulations}
\subsection{Simulation Setup}

In the numerical evaluations we consider an uplink OFDMA MU-MIMO scenario with $U=5$ UEs transmitting towards a single BS, all being active at the considered subcarrier $c$ simultaneously. In addition, there are $V=5$ other UEs which communicate with the BS at the corresponding image subcarrier $c'$. The BS is equipped with an antenna array consisting of $N=20$ antenna elements. Furthermore, each UE has $M_u = M_v = 2$ TX antennas, illustrating a typical UE level capability in modern communications systems. Due to the rather low TX antenna number, each UE transmits only $Q_{u \comma c} = Q_{v \comma c'} = 2$ data streams at each subcarrier and for simplicity the precoding is selected to be a simple one-to-one mapping between the data streams and TX antennas. The input signal-to-noise ratio (SNR) in the individual RX branches is equal to 20~dB. 
Here, we define the SNR as the ratio between the total averaged received signal power originating from all TX data streams of a single user, and the noise power. The transmitted data streams are independent and the total TX power of a single UE is equally shared between its TX branches and data streams. We do realize that the chosen scenario in terms of the number of spatially multiplexed UEs and the number of BS antennas is something that is not necessarily yet feasible in today's systems. However, our intention is to push the capabilities of spatial multiplexing beyond the current systems and to understand, in particular, the sensitivity and limitations imposed by RF circuit imperfections in bigger and bigger antenna array deployments at the advent of massive MIMO~\cite{hoydis_massive_2013, larsson_massive_2014, lu_overview_2014}, with a high number of spatially multiplexed UEs. The above basic scenario, in terms of, e.g., the SNR level and the number of BS antennas, is also varied in the evaluations.

At the desired subcarrier as well as at the image subcarrier, we also consider $L_c = L_{c'} = 8$ external single-antenna interferers with equal powers in the simulation setup. The signal-to-interference ratio (SIR) is defined as the ratio between the total averaged received signal power originating from all TX branches of a single user, and the total received power originating from all external interferers. Note that if the number of the interferers is increased, the power of each individual interferer is decreased in order to obtain the same total SIR. The transmission channels between all TX-RX antenna pairs as well as between all interferer-RX antenna pairs are independent and Rayleigh distributed. 

I/Q imbalance is defined in terms of the image rejection ratio (IRR) given in decibels for a single transceiver branch by $\text{IRR} = 10 \text{log}_{10} (|K_1|^2 / |K_2|^2)$~\cite{anttila_circularity-based_2008}. Firstly, the minimum allowable IRR (IRR$_{\text{min}}$) is set to 25~dB which can be considered to be a typical value for the radio front-end in mobile UEs, as defined, e.g., in 3GPP LTE/LTE-Advanced UE specifications~\cite{3gpp_evolved_2014}. Secondly, we draw phase imbalance coefficients $\phi_{\text{Tx,} u \comma m \comma c}, \forall u, m, c$ and $\phi_{\text{Rx,} c \comma n}, \forall n, c$ independently from $\mathcal{U}(-\alpha, \alpha)$ where $\alpha$ guarantees the selected IRR$_{\text{min}}$ if the gain imbalance was set to zero. Finally, the gain imbalance coefficients $g_{\text{Tx,} u \comma m \comma c}, \forall u, m, c$ and  $g_{\text{Rx,} c \comma n}, \forall n, c$ are selected independently from the conditional distribution $\mathcal{U}(g_{\text{min}}, g_{\text{max}})$ where the range edges correspond to IRR$_{\text{min}}$ with the earlier selected $\phi$. The I/Q imbalance parameters at different subcarriers are assumed to be independent for modeling arbitrarily frequency selective I/Q imbalance. The basic simulation parameters are summarized in Table \ref{parameter_table} while many of the parameters are also systematically varied in the evaluations.

All results describe the performance from a single yet arbitrary subcarrier point of view due to the subcarrier-dependent data streams in the OFDMA systems. In order to illustrate the obtainable performance on average, the results are averaged over all data streams, UEs and 2000 realizations of the I/Q imbalance parameters and the underlying complex fading variables. For each realization, the channel matrices and I/Q imbalance parameters are randomly and independently generated according to the aforementioned criteria. All evaluations are carrier out for both the linear and the augmented linear MMSE RXs. Furthermore, both SINRs and symbol-error rates (SERs) are evaluated.

\begin{table}[t]
	\center
	\caption{Basic simulation parameters}
	\begin{tabular}{ l | l | l}
		\hline \tstruta
		Parameter &Symbol &Value
		\\ \hline \hline \tstrutb
		RX antennas &$N$ &20 \\
		Number of UEs	&$U$, $V$	&5 \\
		TX antennas in UEs &$M_u$, $M_v$ &2 \\
		Data streams in UEs	&$Q_{u \comma c}$, $Q_{v \comma c'}$ &2 \\
		Number of external interferers &$L_c$, $L_{c'}$	&8 \\
		TX antennas in ext. interferers &$J_l$ &1 \\
		Signal to noise ratio	&SNR &20~dB \\
		Signal to interference ratio &SIR$_c$, SIR$_{c'}$	&-20~dB \\
		Minimum image rejection ratio	&IRR$_{\text{min}}$	&25~dB \\
		\hline
	\end{tabular} 
	\label{parameter_table}
\end{table}

\subsection{SINR and SER Simulation Results and Analysis}
\subsubsection{SINR as a function of the SIR}

\begin{figure}[t]
	\centering
	\includegraphics[width=\figurewidth] {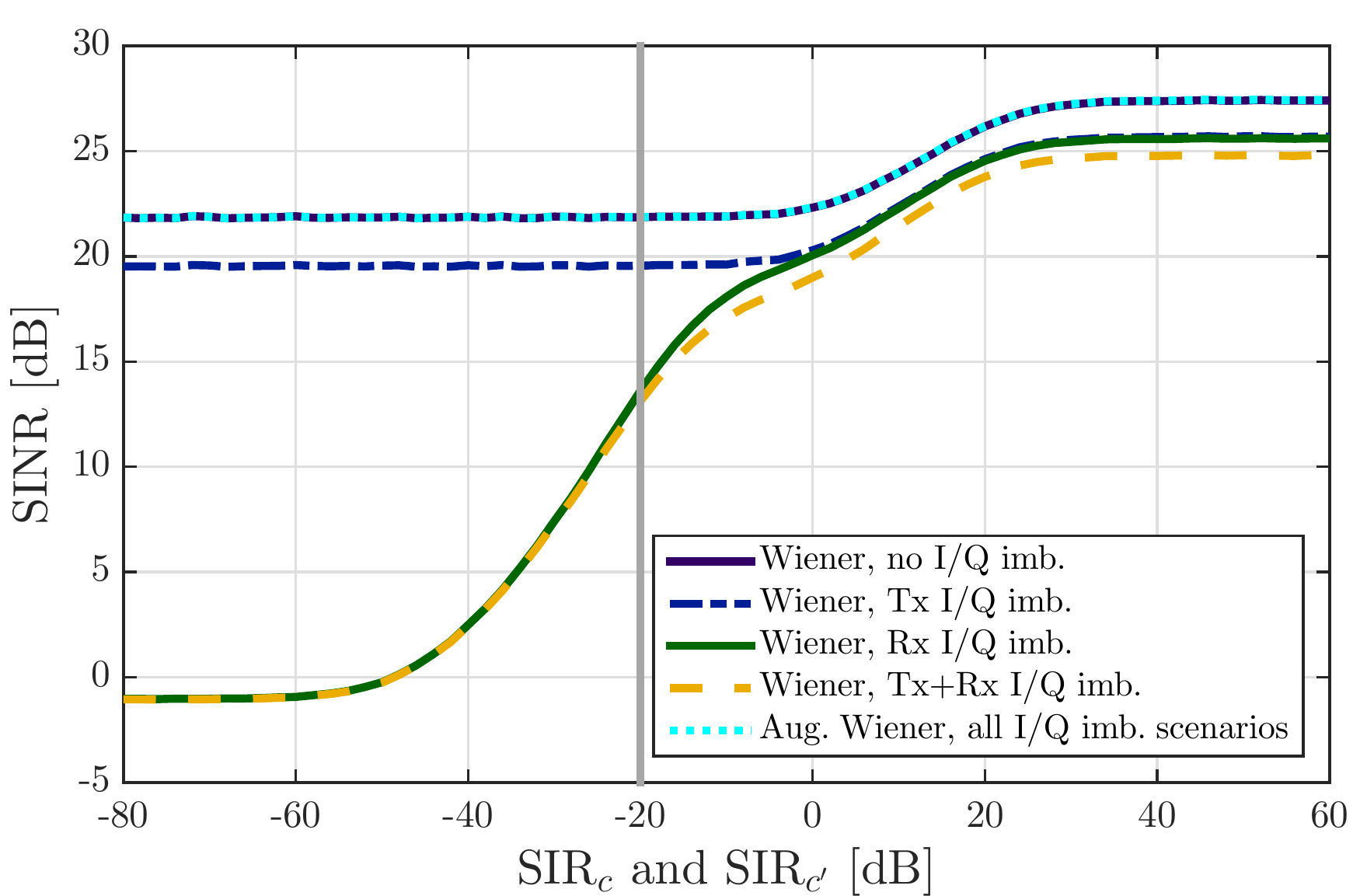}
	\caption{Average SINR as a function of the SIR at both the desired subcarrier and the image subcarrier when the other parameters are fixed. The gray vertical line shows the operating point under the basic conditions given in Table \ref{parameter_table}.}
	\label{SIR_sweep}
\end{figure}

The SINR as a function of the SIR is depicted in Fig.~\ref{SIR_sweep}. Here the power of the external interference is swept at both the desired subcarrier and the image subcarrier while the  useful signal powers are kept equal and constant at both subcarriers. First of all, we notice saturation of the performance with high and low SIRs, even with perfect I/Q matching. In the {\it high SIR region}, the combiners can suppress the inter-user interference effectively and the influence of the external interference is very small. Therefore, the ceiling effect is mainly caused by the additive noise and the spatially multiplexed streams of different UEs. The resulting SINR is actually better than the input SNR since the effect of the noise can be decreased with the antenna array processing, as in the noise limited case, $N=20$ RX antennas provide extra degrees of freedom relative to separating 5$\times$2 = 10 overall streams at the considered subcarrier. Under I/Q imbalances, on the other hand, the performance is limited due to the signal leakage from the image subcarrier UEs. Clearly, TX and RX I/Q imbalances are equally deteriorating the overall performance in high SIR region. The worst SINR is, in turn, seen with joint TX+RX I/Q imbalances which result in an approximately 2.6~dB worse SINR compared to the perfect I/Q matching case. 

When the {\it SIR decreases}, the external interference gets stronger and starts to dominate the received signal. Consequently, the combiners put structurally more effort into the external interference suppression. This is done through the spatial response of the interferers, i.e., the RXs effectively generate nulls to their responses such that the external interference coming through the spatial channels is suppressed efficiently. Actually, the SINR saturation visible in the {\it low SIR region} tells us that a state, where practically all effects of the external interference are suppressed, can be found in each of the cases. However, this kind of interference suppression takes lots of resources. Consequently, the resulting SINR is decreased since the combiners cannot provide as high UE stream separation and noise suppression as with higher SIRs. This is especially visible with the per-subcarrier Wiener combiner under RX and TX+RX I/Q imbalances. In those cases the combiner must mitigate not only the interference from the subcarrier $c$ but also from the mirror subcarrier $c'$ and therefore the combiner easily runs out of degrees of freedom. Under TX I/Q imbalance the situation is easier since the external interference does not leak to subcarrier $c$. The SINR degradation is then caused purely by the mirroring effect on the TX side which causes inter-user interference between the UEs at the mirror subcarriers. The augmented combiner, in turn, provides the same performance under all I/Q imbalance scenarios as the per-subcarrier combiner with ideal I/Q matching. In these cases the influence of the external interference and the inter-user interference as well can be suppressed very efficiently while still providing a slight array processing gain (the original SNR is 20~dB). 
We conclude that, in general, and as obvious in Fig.~\ref{SIR_sweep}, the augmented combiner can provide substantial enhancement in the output SINR, especially under high levels of the external interference.

\subsubsection{SINR as a function of the SIR$_{c'}$}

\begin{figure}[t]
	\centering
	\includegraphics[width=\figurewidth] {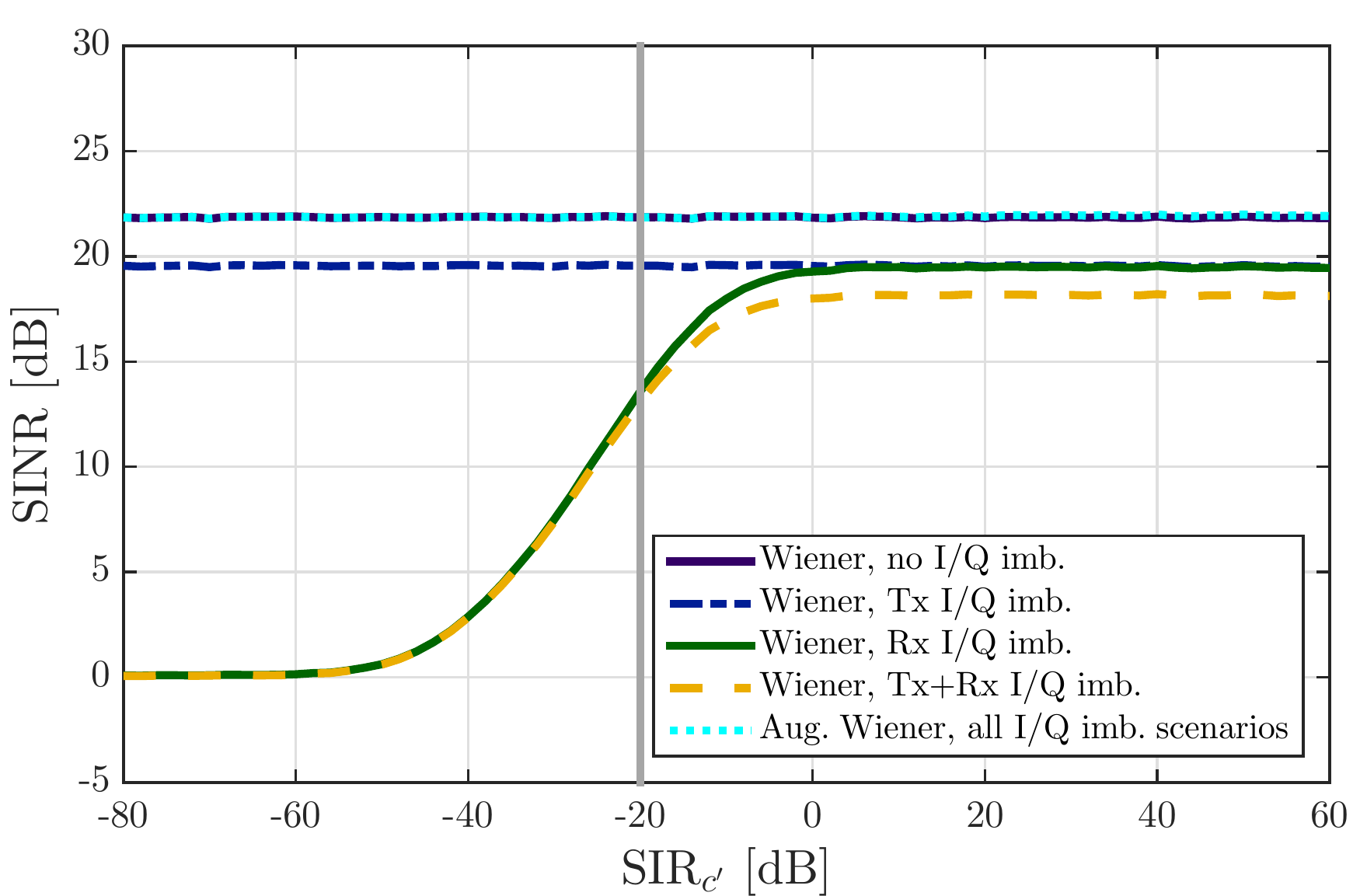}
	\caption{Average SINR as a function of the SIR at the image subcarrier when the other parameters are fixed. The gray vertical line shows the operating point under the basic conditions given in Table \ref{parameter_table}.}
	\label{SIR_c_p_sweep}
\end{figure}

The leakage of the external interference at the image subcarrier is further illustrated in Fig. \ref{SIR_c_p_sweep}. It shows the SINR when the SIR at the image subcarrier is swept while the SIR at the desired subcarrier is fixed to -20~dB. Based on the results, the augmented combiner has a flat and robust response over all SIR$_{c'}$ values and with all I/Q imbalance scenarios. This means that it can suppress the effect of the signal leakage very efficiently and thus provides good performance in all conditions. Also the per-subcarrier processing under TX I/Q imbalance only yields a flat response which has 2.3~dB lower SINR level than with the augmented combiner. This difference is purely caused by the inter-user interference from the image subcarrier since the external interference and noise at the image subcarrier are not affected by TX I/Q imbalance. The response of the ordinary Wiener combiner is not flat when considering I/Q imbalance in the RX side. In that case, the SINR drops drastically as SIR$_{c'}$ decreases. When comparing Figs. \ref{SIR_sweep} and \ref{SIR_c_p_sweep} with each other, we notice that actually the interference leakage is the main reason for performance degradation with high external interference levels since in those cases the resulting SINRs are almost the same in both figures. 
Also now, the combiner puts most of its structural resources to interference suppression and reaches saturation of the SINR. However, it simultaneously sacrifices lots of the data stream separation and noise suppression capabilities and consequently the resulting SINR level is fairly low. The results clearly indicate that the overall SINR performance of the ordinary per-subcarrier Wiener combiner is heavily deteriorated by the strong external interference at the image subcarrier, even if the contribution of the external interference at the considered subcarrier can be efficiently suppressed. 

\subsubsection{SINR as a function of the SNR}

\begin{figure}[t]
	\centering
	\includegraphics[width=\figurewidth] {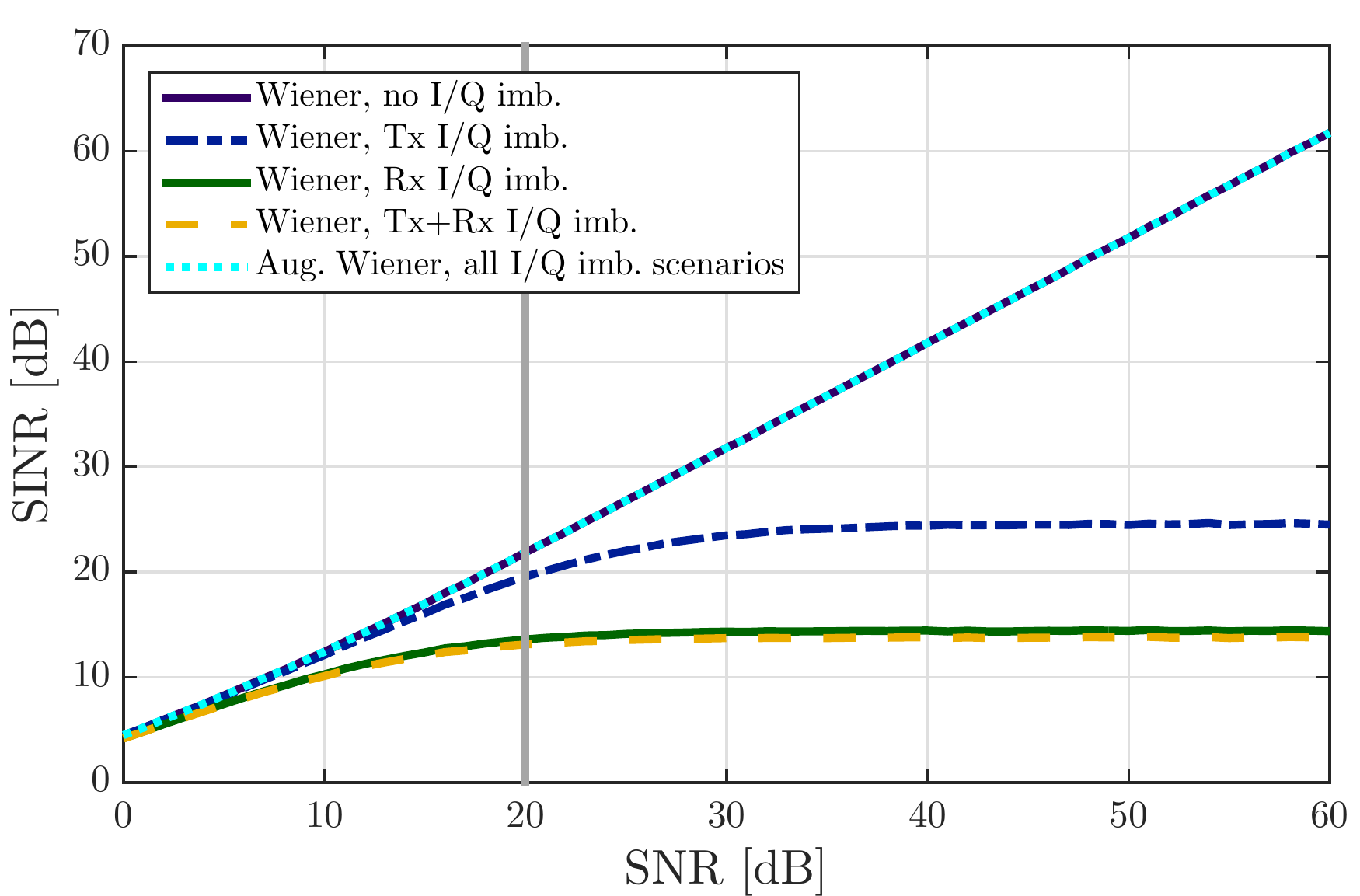}
	\caption{Average SINR as a function of the SNR when the other parameters are fixed. The gray vertical line shows the operating point under the basic conditions given in Table \ref{parameter_table}.}
	\label{SNR_sweep}
\end{figure}

Fig. \ref{SNR_sweep} visualizes the SINR as a function of the input SNR. The performance saturates under I/Q imbalances and the worst performance with the Wiener combiner is obtained if I/Q imbalance occurs in the RX electronics. The ceiling effect, due to the unavoidable signal leakage from the image subcarrier, is very strong and the SINR saturates at around 25~dB SNR with RX and TX+RX imbalance scenarios and at around 35~dB SNR with TX I/Q imbalance. At these points, the external interference and inter-user interference, both from subcarriers $c$ and $c'$, fully dominate the SINR behavior and the contribution of the noise is practically negligible. Again, the augmented combiner outperforms the conventional one clearly and results in a linear growth of the SINR against input SNR. The results in Fig. \ref{SNR_sweep} also extend the work related to the SU-SIMO scenario in~\cite{hakkarainen_interference_2014} and show somewhat similar behavior in both cases. 

\subsubsection{SINR as a function of L}

\begin{figure}[t]
	\centering
	\includegraphics[width=\figurewidth] {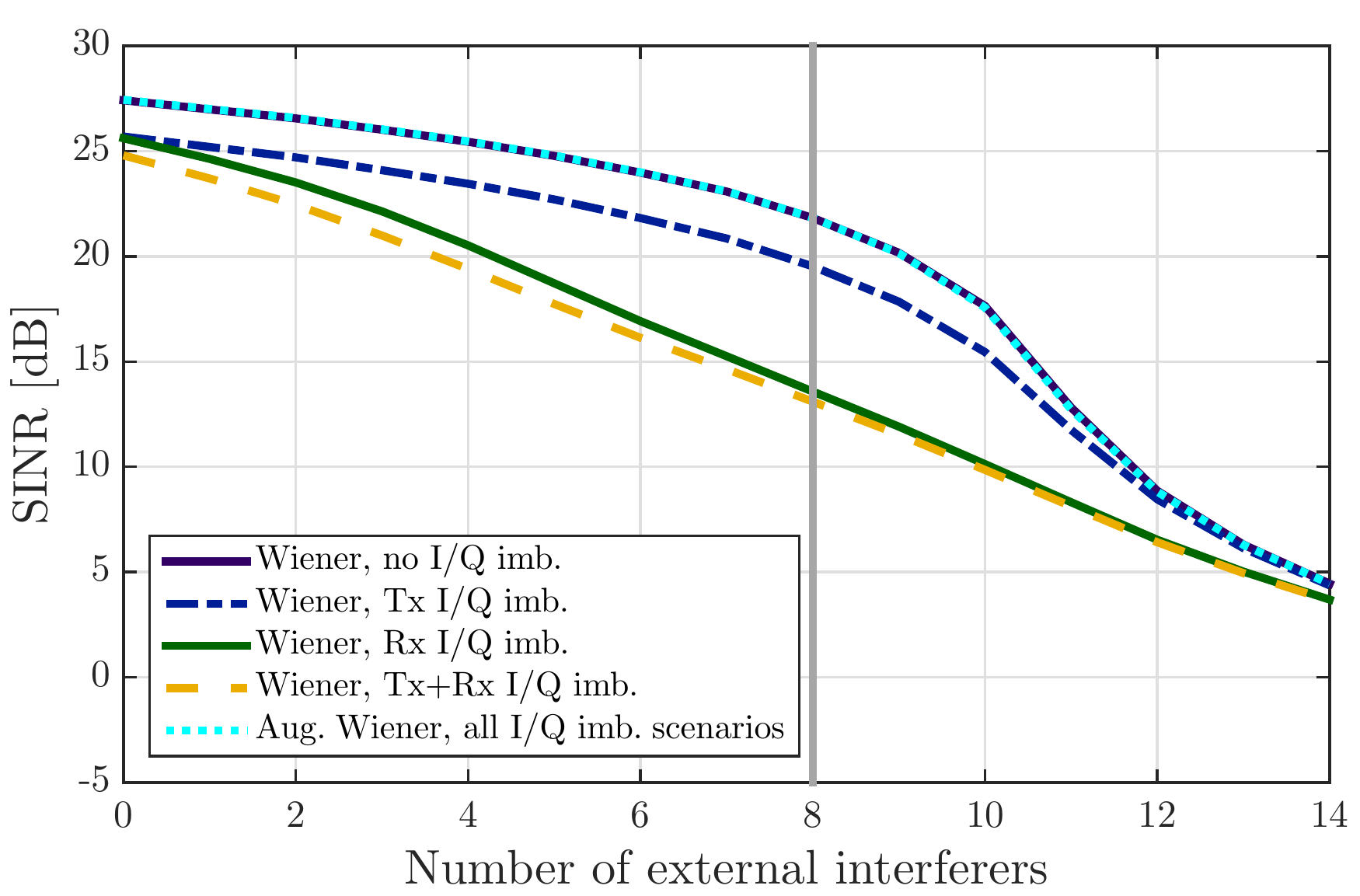}
	\caption{Average SINR as a function of the number of external interferers $L$ when the other parameters are fixed. The gray vertical line shows the operating point under the basic conditions given in Table \ref{parameter_table}.}
	\label{L_sweep}
\end{figure}

The effect of increasing the number of external interferers is depicted in \mbox{Fig. \ref{L_sweep}}. With the simulation parameters given in Table \ref{parameter_table}, there are \mbox{$M_u U+ J_l L = 18$} incoming signals at the desired subcarrier as well as at the image subcarrier. In theory, the linear combiners are able to separate $N=20$ signals as long as all the signal sources have separable spatial characteristics, i.e., their channel responses are not fully correlated. Thus the number of single-antenna interferers could be even increased to $L=10$, resulting in 20 incoming signals in total, without losing the ability for signal separation in theory. However, based on the figure, the SINR decreases as the number of interferers increases, even for less than $L=10$ external interferers. This is natural as optimizing the MSE at the combiner output, corresponds to finding a proper compromise between coherent combining of the desired signal as well as suppressing the inter-stream, inter-user and external interference as well as noise, and all of their mirror images. Thus, when the number of signals increases, the optimization task becomes increasingly difficult. 
The best SINR is provided by the augmented combiner under any I/Q imbalance scenario whereas the per-subcarrier Wiener processing under RX and TX+RX I/Q imbalances turns out to have the worst SINRs. This is again caused by the interference leakage from the image subcarrier and is now emphasized since the number of the interferers is swept at both subcarriers. When the number of interferers exceeds 10, also the augmented combiner runs out of degrees of freedom in interference suppression and consequently the SINRs of all scenarios drop steeply towards lower levels. 

\subsubsection{SINR as a function of the IRR$_{\textnormal{min}}$}

\begin{figure}[t]
	\centering
	\includegraphics[width=\figurewidth] {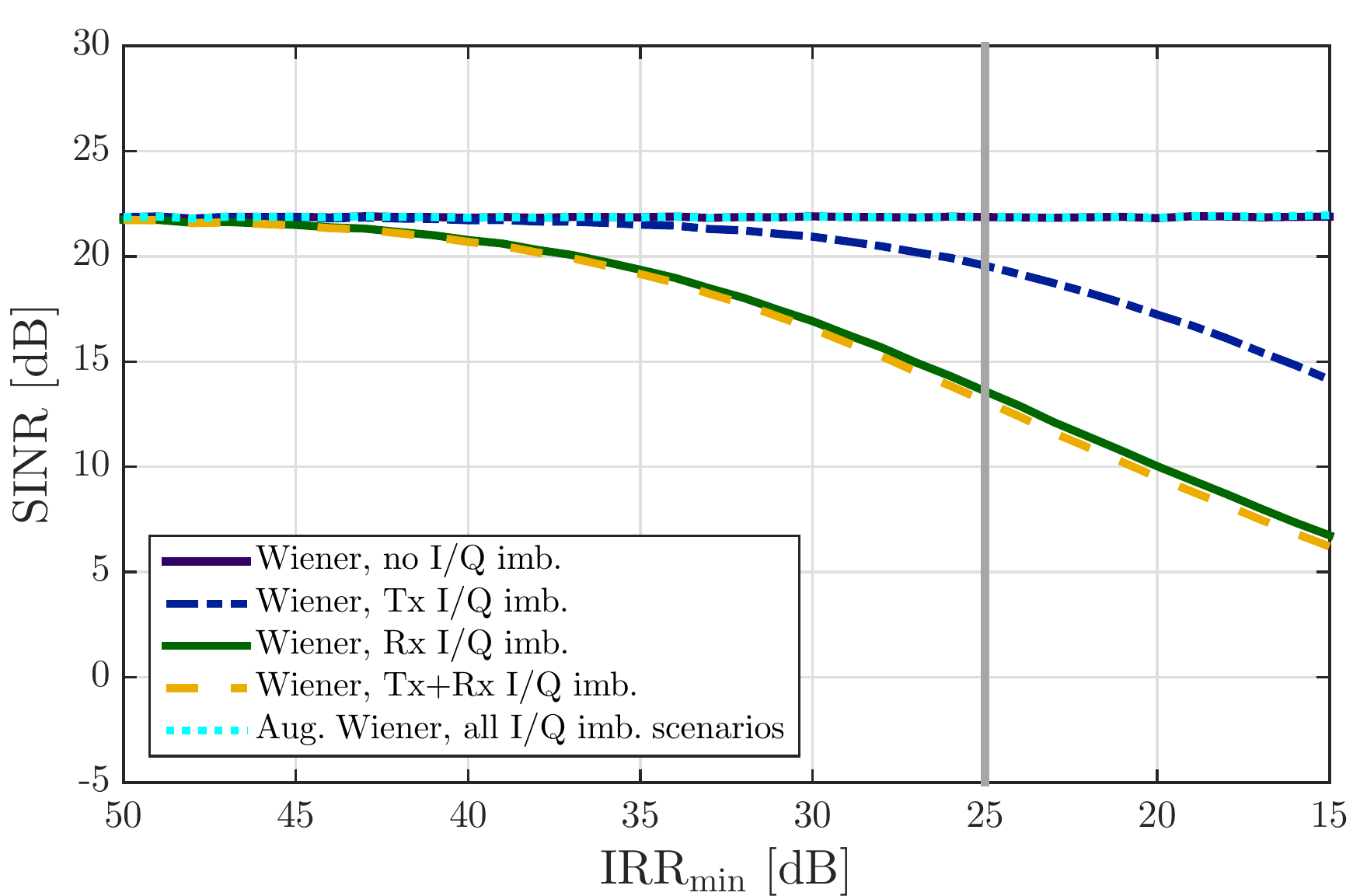}
	\caption{Average SINR as a function of the IRR$_{\text{min}}$ when the other parameters are fixed. The gray vertical line shows the operating point under the basic conditions given in Table \ref{parameter_table}.}
	\label{IRR_sweep}
\end{figure}

Fig. \ref{IRR_sweep} shows the SINR performance when the minimum allowable IRR is varied. The augmented combiner produces a flat response for all IRR$_{\text{min}}$ values, meaning that the effects of I/Q imbalances are mitigated completely even for low IRRs. The performance of the ordinary per-subcarrier processing under TX I/Q imbalance is deteriorated by the inter-user interference from the image subcarrier and therefore the SINR degrades fairly slowly as IRR$_{\text{min}}$ decreases. In contrast, the SINR under RX I/Q imbalance is heavily degraded, again due to the increasing external interference leakage from the image subcarrier. It is worth noting that under RX or TX+RX I/Q imbalances and even with very moderate values of IRR$_{\text{min}}$ the SINR is degraded by several decibels. Even with IRR$_{\text{min}} = 35$~dB, which is already a challenging number to achieve systematically, the per-subcarrier Wiener processing is some 2.5~dB below the SINR of the augmented Wiener combiner.

\subsubsection{SINR as a function of N}

\begin{figure}[t]
	\centering
	\includegraphics[width=\figurewidth]{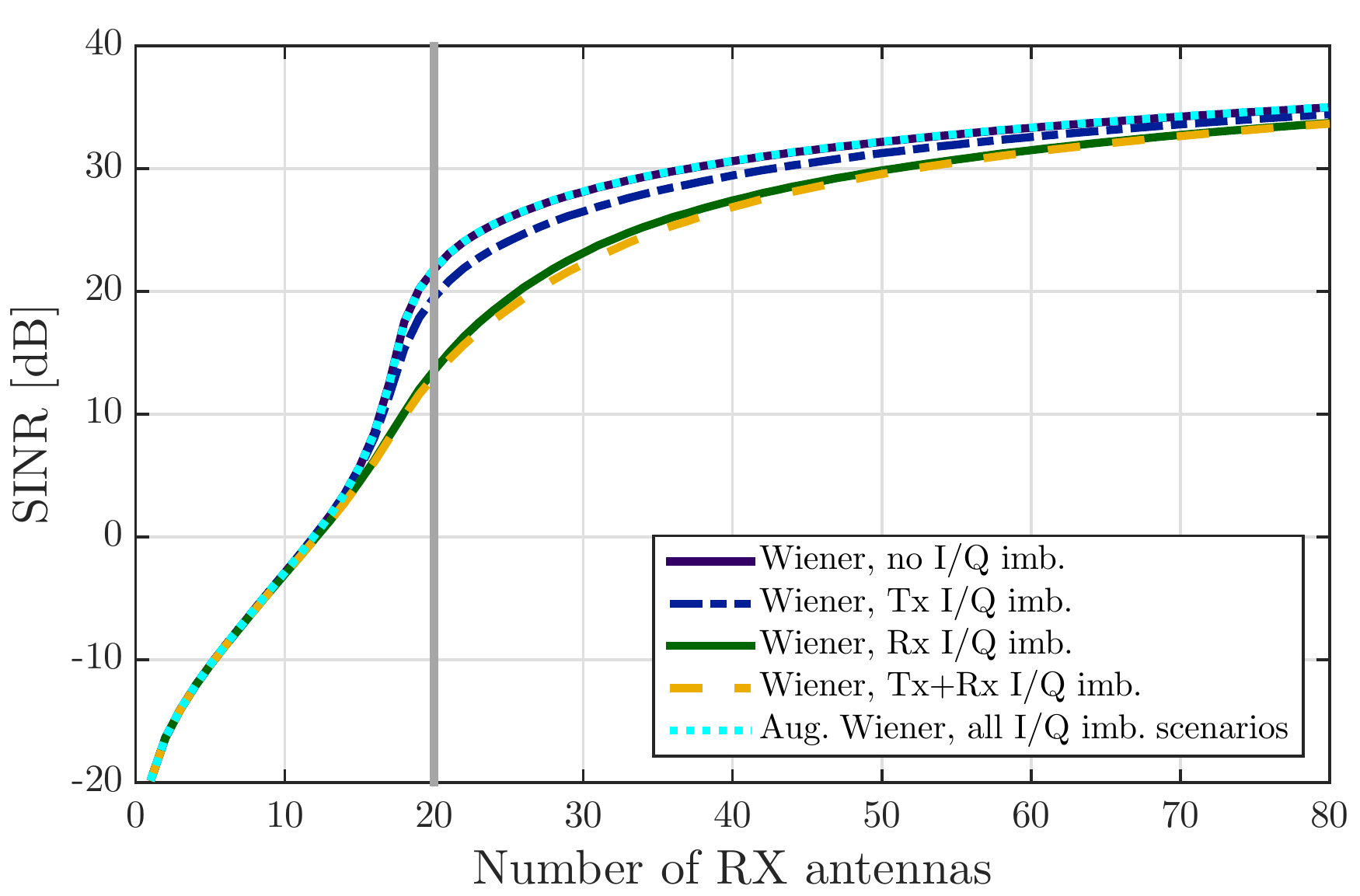}
	\caption{Average SINR as a function of the number of RX antennas $N$ when the other parameters are fixed. The gray vertical line shows the operating point under the basic conditions given in Table \ref{parameter_table}.}
	\label{N_sweep}
\end{figure}

Fig. \ref{N_sweep} illustrates the SINR as a function of the number of RX antennas.  Although modern communications systems usually support at most 4--8 antennas to be used in the BSs, this figure shows the capability of antenna array processing and thus gives an important insight also for the behavior towards emerging massive MIMO systems under I/Q imbalances, see e.g.~\cite{hakkarainen_widely-linear_2013}.
Based on the results with varying number of RX antennas, the performance is really poor when the number of RX antennas is around 12 or less due to too little degrees of freedom to spatially separate the signals. Beyond that point, the RX starts to be able to separate different signals and the SINR of the augmented combiner grows very steeply as $N$ increases. Also the per-subcarrier Wiener processing under TX I/Q imbalance only gets a similar performance boost. The both curves start to saturate after the point where the number of antennas matches with the number of incoming signals which is in this case equal to \mbox{$M_u U+ J_l L = 18$}. In contrast to these curves, RX and TX+RX I/Q imbalances cause slower increase in the resulting SINR and their saturation starts later, around the point $N = 28$. That point coincides with $N = M_u U + 2 J_l L$ which means that at this point the per-subcarrier Wiener processing is finally able to separate the signals from the desired subcarrier and strong interferers at both subcarriers from each others. Thus it is able to provide the same SINR as the augmented combiner has already with $N=20$ antennas. 
As the number of antennas becomes very high, both combiners perform well under all I/Q imbalance scenarios. Additionally, the SINR increases only slightly when adding RX antennas to the BS side. This is a consequence of the situation where both combiners have more than enough spatial resources and they can use the extra degrees of freedom purely for noise optimization and interference suppression purposes. 

\subsubsection{SER as a function of SNR}

\begin{figure}[t]
	\centering
	\includegraphics[width=\figurewidth]{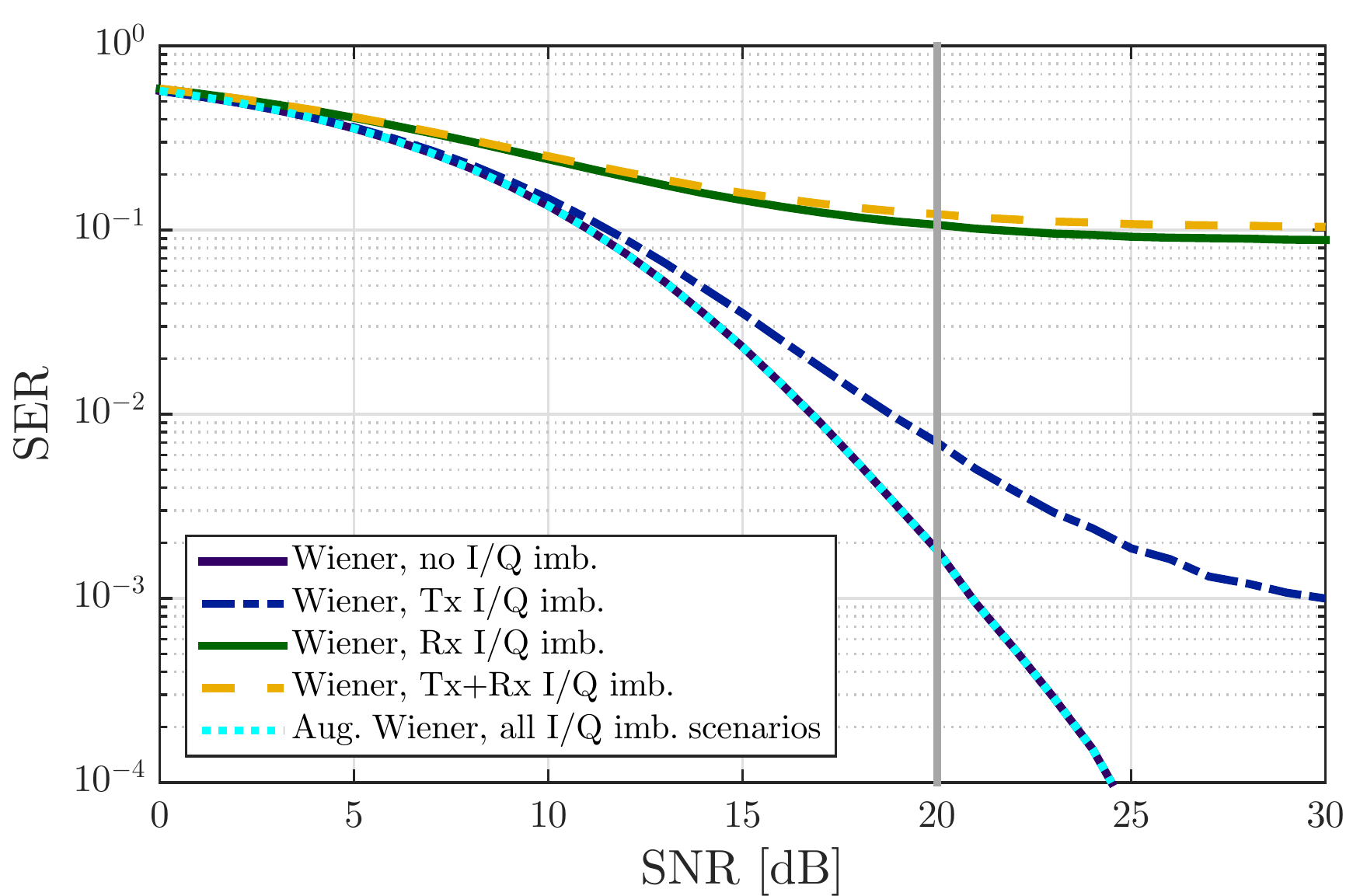}
	\caption{Average uncoded SER as a function of the SNR when the other parameters are fixed and 16-QAM signal constellation is used. The gray vertical line shows the operating point under the basic conditions given in Table \ref{parameter_table}.}
	\label{SER_SNR}
\end{figure}

In order to evaluate the performance of the augmented Wiener combiner with respect to another commonly used metric we next provide the uncoded SER performance as a function of the SNR in Fig.~\ref{SER_SNR}. Here, we use 16-QAM signal constellation and all SERs are averaged over all data streams of all UEs and over 20000 realizations of the I/Q imbalance parameters and the underlying complex fading variables.
We notice that the SER decreases as the SNR increases in all cases, which is of course natural. Additionally, we see that RX I/Q imbalance with the per-subcarrier combiner causes the SER to saturate to high levels. This was also expected based on the SINR results given in Fig.~\ref{SNR_sweep}. Note that the exact level of the saturation might vary due to different antenna setups, I/Q imbalance parameters, signal constellations, channel models etc. Similar to RX I/Q imbalance only, saturation of the SER is visible also with joint TX+RX I/Q imbalances, again with the per-subcarrier combiner. The level of the saturation is even slightly higher than with RX I/Q imbalance only. When considering I/Q imbalance only on the TX side, the SER performance is much better than under RX and TX+RX I/Q imbalances. In addition, such a strong SER saturation is not visible under TX I/Q imbalance. The augmented combiner, in turn, provides the same SER as a system under ideal I/Q matching and consequently, no performance degradation is seen due to any I/Q imbalance. Note that this is a significant difference to the per-subcarrier combiner which suffers heavily from the signal distortion.

\subsubsection{SER as a function of ${\text{IRR}}_{\textnormal{min}}$}

\begin{figure}[t]
	\centering
	\includegraphics[width=\figurewidth]{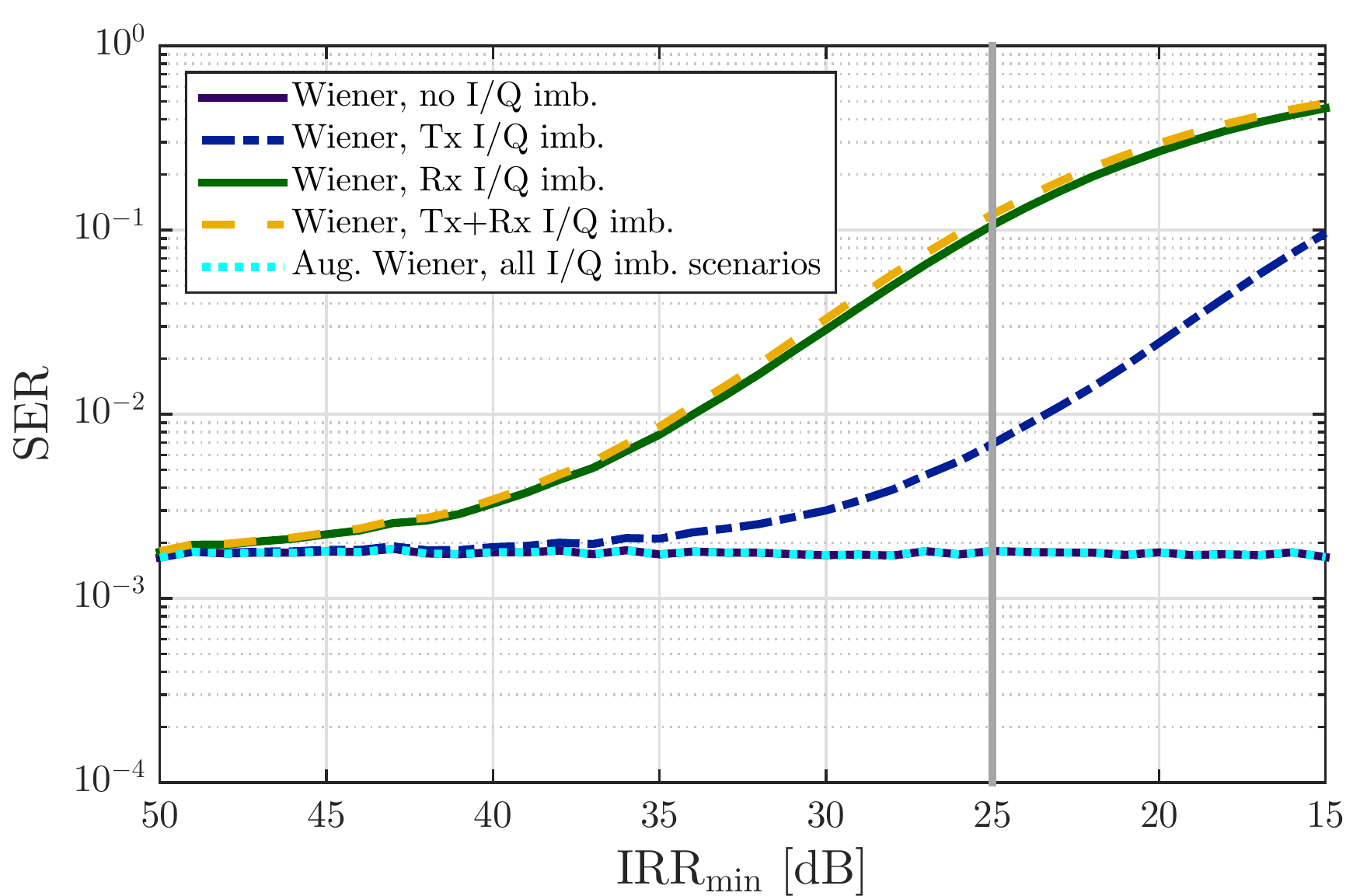}
	\caption{Average uncoded SER as a function of the IRR$_{\text{min}}$ when the other parameters are fixed and 16-QAM signal constellation is used. The gray vertical line shows the operating point under the basic conditions given in Table \ref{parameter_table}.}
	\label{SER_IRR}
\end{figure}

The performance evaluation is continued in Fig.~\ref{SER_IRR} where the SER is depicted as a function of the ${\text{IRR}}_{\text{min}}$. As seen already in the SINR evaluations in Fig.~\ref{IRR_sweep}, the performance degrades as the ${\text{IRR}}_{\text{min}}$ decreases. The SER evaluations clearly indicate that the augmented combiner structure can very effectively suppress the effects of I/Q imbalance and thus provides new possibilities, e.g., in terms of cheaper RF components or more robust operation. In contrast, the SER of the conventional per-subcarrier combiner is highly deteriorated by I/Q imbalances and consequently, the system performance is degraded. 

\subsection{Further Aspects in Massive MIMO Framework}

In order to get the massive MIMO concept into reality, research and profound understanding are needed regarding the associated RF imperfections~\cite{lu_overview_2014, larsson_massive_2014}. It is indicated in some studies in the existing literature that massive MIMO systems are robust to RF imperfections or the effect of the imperfections is very small, see e.g.~\cite{bjornson_massive_HW_2014, bjornson_massive_2014}. This statement may hold for simple cases where the effect of RF impairments is modeled as additive uncorrelated Gaussian noise. However, in general, RF impairments distort the transmitted and received signals in a more complicated manner and the resulting signal distortion is dependent, e.g., on the signal power and subcarrier allocation scheme, and consequently the results with simplified distortion models may not be precise or valid anymore.

In this subsection, we specifically focus on practical aspects of I/Q imbalances in massive MIMO deployments, building on our earlier signal and system modeling, and consider two specific scenarios. Toward this end, we modify the system scenario and parameterization compared to what we had in the previous subsection. In particular, we increase the number of RX antennas considerably, being eventually an order of magnitude higher than the number of spatially multiplexed UEs at each subcarrier which is a typical assumption in massive MIMO systems~\cite{hoydis_massive_2013, larsson_massive_2014, bjornson_massive_2014}. As also in the previous subsection, we assume that each RX antenna is connected to a separate RX branch. Moreover, we assume that the UEs are simple single-antenna devices, i.e., $M_u = 1$, and consequently set also the number of transmitted data streams in each UE to $Q_{u \comma c} = 1$. This way the network is considered to support also low-cost and simple UEs which are, e.g., a crucial element in the increasingly popular internet of things (IoT) concept. We also fix $\text{IRR}=20$~dB since the transceivers in massive MIMO systems, especially in the BS side, are considered to be implemented with low-cost components which, in turn, are prone to severe impairments~\cite{larsson_massive_2014, bjornson_circuit-aware_2014, bjornson_massive_2013}. In addition, we assume no external interference, i.e., $L=0$,  since the massive MIMO concept is often considered to be adopted at centimeter or millimeter wave frequencies where the interference even from closely located devices may be low due to high propagation losses. Finally, there has been some speculation that, in order to achieve a very simple system, massive MIMO could be adopted without frequency multiplexing through the OFDMA principle, thus resulting in a plain OFDM based scheme where simultaneous UE multiplexing is carried out only spatially. Therefore, in the following, we set $u=v$ and $U=V$, i.e., the same set of UEs use both subcarriers $c$ and $c'$. All parameters for the massive MIMO scenario are summarized in Table \ref{tab:mm_parameters}. 

\begin{table}[t]
	\center
	\caption{Basic simulation parameters for massive MIMO setup.}
	\begin{tabular}{ l | l | l}		
		\hline \tstruta
		Parameter &Symbol &Value
		\\ \hline \hline \tstrutb
		RX antennas &$N$ &100 \\
		Number of UEs &$U$ &5 \\
		TX antennas in UEs &$M_u$	&1 \\
		Data streams in UEs	&$Q_{u \comma c}$ &1 \\
		Number of external interferers &$L_c$, $L_{c'}$ &0 \\
		Signal to noise ratio	&SNR &20~dB \\
		Image rejection ratio	&IRR &20~dB \\
		\hline
	\end{tabular} 
	\label{tab:mm_parameters}
\end{table}

Massive MIMO systems require extreme simplicity not only in the hardware but also in the associated signal processing~\cite{larsson_massive_2014}. Consequently, we adopt maximum ratio combining (MRC), which is known of its low computational burden and straightforward implementation, as a benchmark against the per-subcarrier Wiener and augmented Wiener approaches. In general, the classical MRC weights for a single UE are of the form $\mathbf{W}_{u \comma c}^{\text{MRC}} = \mathbf{H}_{u \comma c}$~\cite{ngo_energy_2013}. However, under RF impairments, the effective channel includes also the influence of imperfect electronics in the TX and RX and consequently, stemming from our earlier modeling, the MRC weights under I/Q imbalances become equal to 
\begin{align}
	\mathbf{W}_{u \comma c}^{\text{MRC}} = \widetilde{\mathbf{\Psi}}_{u \comma c}.
	\label{eq:w_MRC}
\end{align}
Here, MRC is, indeed, assumed to be aware of the user-specific effective propagation channel $\widetilde{\mathbf{\Psi}}_{u \comma c}$ incorporating partially the I/Q imbalance response as given in (\ref{eq:PsiOmega}). This assumption is justified since in practice the channel estimation is really done for the effective spatial channel matrix and thus it does include also the effects of non-ideal transceivers. It is worth noting that the MRC detector of UE $u$ can utilize neither the channel information of the other UEs nor of the possible external interferers. 

\begin{figure}[t]
	\centering
	\includegraphics[width=\figurewidth]{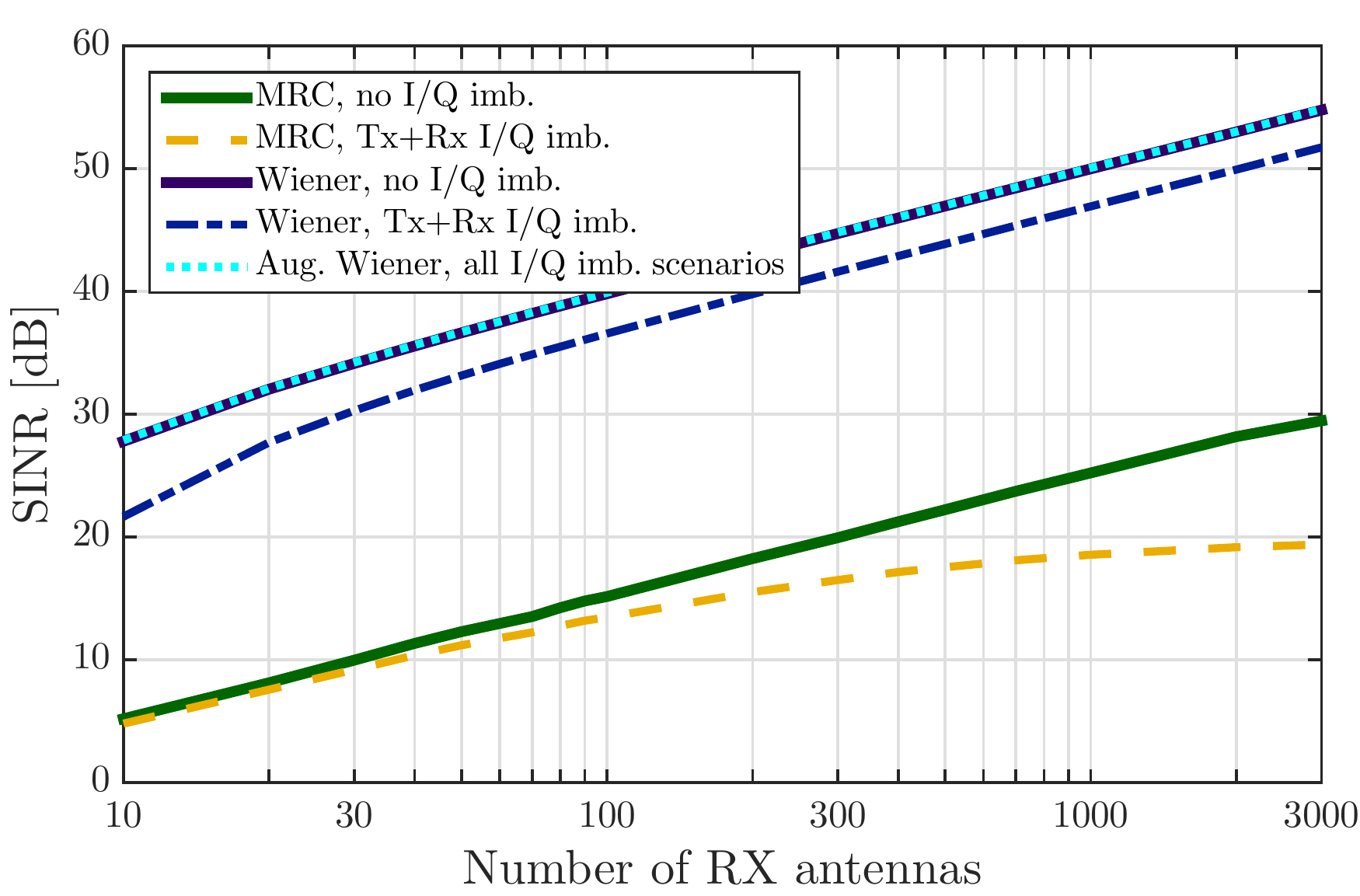}
	\caption{Average SINR as a function of the number of RX antennas $N$ for massive MIMO scenario. Note the logarithmic x-axis. Basic simulation parameters are given in Table \ref{tab:mm_parameters}.}
	\label{N_sweep_MM}
\end{figure}

In the numerical evaluations, we focus now on the influence of I/Q imbalances while varying the number of RX antennas and the number of UEs. Fig. \ref{N_sweep_MM} illustrates the SINR of the MRC, Wiener and augmented Wiener approaches as a function of the number of RX antennas. At first, we notice that all methods have a slope equal to $10\text{log}_{10}(N)$ under perfect I/Q matching, i.e., without I/Q imbalance. What is interesting, however, is that MRC has some 25~dB worse SINR than the other methods. This is caused by the fact that MRC cannot structurally suppress any interference including, in this case, also the inter-user interference from the other spatially multiplexed UEs. This also means that, under the considered scenario, MRC requires roughly 300 times more RX antennas than Wiener and the augmented Wiener in order to provide an equal SINR performance, which is practically not feasible. Under I/Q imbalances, the SINR performance is even more interesting. The SINR of MRC does not anymore follow the slope of $10\text{log}_{10}(N)$. In fact, it saturates to 20~dB {\it even when the number of RX antennas approaches infinity}. This is explained by the following fact. As visible in (\ref{eq:w_MRC}), the MRC weights are matched to the effective channel $\widetilde{\mathbf{\Psi}}_{u \comma c}$ which, in turn, is dominated by the term $\mathbf{K}_{\text{Rx1,} c} \mathbf{H}_{u \text{,} c} {K}_{\text{Tx1,} u \text{,} c}$, as given in (\ref{eq:PsiOmega}). However, when interpreting the received signal (\ref{r_TxRxi}) from the OFDM perspective, the inter-carrier interference from the same UE propagates through $\widetilde{\mathbf{\Omega}}_{u \comma c}$ which includes a term $\mathbf{K}_{\text{Rx1,} c} \mathbf{H}_{u \text{,} c} {K}_{\text{Tx2,} u \text{,} c}$, see again (\ref{eq:PsiOmega}). Thus, the only difference lies in the different TX I/Q imbalance scaling factors. Based on this, we conclude that \emph{the SINR of MRC under TX+RX I/Q imbalances is limited to} $10\text{log}_{10}(|{K}_{\text{Tx1,} u \text{,} c}|^2 / |{K}_{\text{Tx2,} u \text{,} c}|^2 )$ \emph{which is exactly the same as the TX IRR}. In contrast, the per-subcarrier Wiener has 3--6~dB loss in the SINR compared to the no I/Q imbalance case. This is caused purely by the leakage of the UE signals from the image sucarrier since there are no external interference sources involved now. It is, however, important to note that the Wiener method still provides the same slope in the SINR and thus its performance is not restricted to any fixed upper bound. The augmented Wiener under TX+RX I/Q imbalance has, again, an equal performance to a system with no I/Q imbalance and hence it outperforms the per-subcarrier methods clearly.  
For any given SINR target, by using the augmented Wiener processing, one can thus lower the number of deployed antennas or the transceiver I/Q matching specifications, or both.

\begin{figure}[t]
	\centering
	\includegraphics[width=\figurewidth]{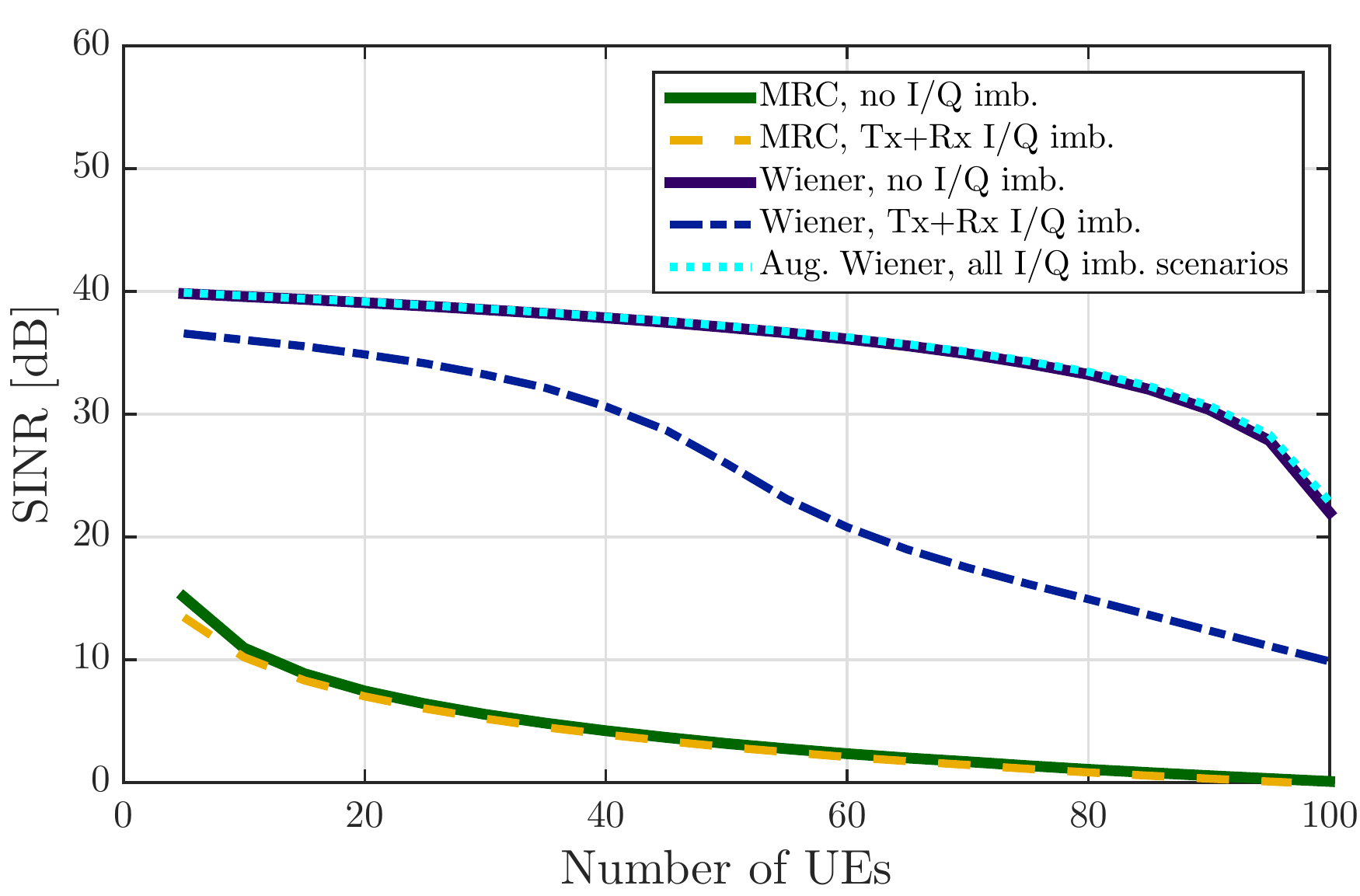}
	\caption{Average SINR as a function of the number of UEs $U$ for massive MIMO scenario. Basic simulation parameters are given in Table \ref{tab:mm_parameters}.}
	\label{U_sweep_MM}
\end{figure}

Next, the SINR as a function of the number of UEs is depicted in Fig. \ref{U_sweep_MM}. As expected, the increasing number of spatially multiplexed UEs decreases the SINR in all cases. Naturally, this stems from the increased inter-user interference as well as from the limited degrees of freedom in the RX processing. Also now, MRC has the worst SINR performance and this time there are no big differences between the I/Q imbalance scenarios. It is noticeable that the SINR of MRC may be, in practice, too low for many communications applications especially when the ratio between the number of RX antennas and UEs, i.e. $N/U$, decreases. On the contrary, the per-subcarrier Wiener has much better SINR than MRC. However, it cannot structurally suppress the inter-carrier and inter-user interference and thus suffers heavily from I/Q imbalance. Therefore, the augmented subcarrier processing turns out to have, once more, the best SINR performance among the considered processing methods. Naturally, this performance improvement comes at a cost of more complex combining process but may still provide the best cost/quality ratio even in massive MIMO systems. Furthermore, we want to emphasize that, as discussed in Section III.D, the overall complexity of the whole digital signal processing chain is not necessarily increased dramatically when changing the per-subcarrier processing to the augmented one. 

\section{Conclusion}

Radio transceiver I/Q imbalances in MIMO communications with OFDM waveforms have been widely studied in the existing literature. This paper, however, extended the system approach to multiuser OFDMA-based MIMO uplink where multiple UEs are active simultaneously at each subcarrier and in addition to that frequency division multiplexing is deployed. We also included the effects of possible external interference in the modeling and analysis and thus provided valuable insight for the future heterogeneous network designs where system coexistence and interference suppression are key issues. It was explicitly shown that I/Q imbalances of UE transmitters and BS receiver distort the signal properties and cause inter-carrier and inter-user interference originating from the image subcarrier. This phenomenon turns out to be especially harmful when the external interference at the image subcarrier is strong. Furthermore, I/Q imbalance complicates separating the spatially multiplexed UEs at a given considered subcarrier.

The provided extensive SINR analysis, as a function of multiple system parameters, shows that the performance of the conventional per-subcarrier processing is heavily limited under I/Q imbalances and external interference. Stemming from that, an augmented spatial combiner was formulated, combining the signals jointly between image subcarriers and across all RX antennas. The proposed augmented subcarrier processing mitigates the effects of the transceiver I/Q imbalances efficiently and indeed provides combiner output SINRs practically identical to a reference system with I/Q imbalance free transceivers. Note that the augmented processing is implemented completely by digital signal processing in the BS RX. Thus the number of costly RF chains and demanding FFT processing blocks are equal to those of the conventional per-subcarrier processing and, in fact, we showed that the increase in the computational complexity can be only a few tens of percents when utilizing the augmented processing instead of the conventional one. Moreover, the augmented processing integrates the data stream separation, interference suppression, noise suppression and I/Q imbalance mitigation all into a single processing stage, thus avoiding separate transceiver calibration. The augmented approach was shown to operate very effectively and reliably also in massive MIMO framework whereas the per-subcarrier based processing approaches suffer from limited performance, in spite of the huge number of RX antennas. Overall the results demonstrate that reliable and high-performance spatial processing characteristics can be obtained by the proposed augmented combiner principle, in spite of challenging levels of the external interference, transceiver I/Q imbalances and high number of spatially multiplexed users, in the considered OFDMA MU-MIMO systems.

\section*{Appendix A}
\label{app1}

The power terms in (\ref{P_WL_separated}) can be expressed easily, since the covariance matrix $\widetilde{\mathbf{R}}_{\text{TxRxi,} c}$ in (\ref{R_WL_TxRxi}) is a sum of multiple independent terms. We only need to define two stream selection matrices: $\mathbf{\Gamma}_{q \comma u \comma c} = \text{diag} (\mathbf{e}_q) \in {\mathbb{R}}^{Q_{u \comma c} \times Q_{u \comma c}}$ and \mbox{$\mathbf{\Delta}_{q \comma u \comma c} = \mathbf{I} - \mathbf{\Gamma}_{q \comma u \comma c} \in {\mathbb{R}}^{Q_{u \comma c} \times Q_{u \comma c}}$} which refer to data stream $q$ of user $u$ at subcarrier $c$, and to the interfering other streams of the same UE, respectively. Then the power terms in (\ref{P_WL_separated}) are given with the help of (\ref{R_WL_TxRxi}) and the stream selection matrices by
\begin{align}
	&\widetilde{P}_{x \comma q \comma u \comma c} = 
	\sigma_{\text{x,} u \comma c}^2 \widetilde{\mathbf{w}}_{q \comma u \comma c}\HH \widetilde{\mathbf{\Xi}}_{u \comma c} \mathbf{G}_{u \comma c} \mathbf{\Gamma}_{q \comma u \comma c} \mathbf{G}_{u \comma c}\HH \widetilde{\mathbf{\Xi}}_{u \comma c}\HH \widetilde{\mathbf{w}}_{q \comma u \comma c}
	\label{eq:power_terms1}
\end{align}
\begin{align}
	&\widetilde{P}_{\text{ISI} \comma q \comma u \comma c} =
	\sigma_{\text{x,} u \comma c}^2 \widetilde{\mathbf{w}}_{q \comma u \comma c}\HH \widetilde{\mathbf{\Xi}}_{u \comma c} \mathbf{G}_{u \comma c} \mathbf{\Delta}_{q \comma u \comma c} \mathbf{G}_{u \comma c}\HH \widetilde{\mathbf{\Xi}}_{u \comma c}\HH \widetilde{\mathbf{w}}_{q \comma u \comma c}
	\label{eq:power_terms2}
\end{align}
\begin{align}
	&\widetilde{P}_{\text{IUI} \comma u \comma c} = \sum_{i=1, i \neq u}^{U}
	\sigma_{\text{x,} i \comma c}^2 \widetilde{\mathbf{w}}_{q \comma u \comma c}\HH \widetilde{\mathbf{\Xi}}_{i \comma c} \mathbf{G}_{i \comma c} \mathbf{G}_{i \comma c}\HH \widetilde{\mathbf{\Xi}}_{i \comma c}\HH \widetilde{\mathbf{w}}_{q \comma u \comma c}
	\label{eq:power_terms3}
\end{align}
\begin{align}
	&\widetilde{P}_{\text{IUI} \comma c'} = \sum_{v=1}^{V}
	\sigma_{\text{x,} v \comma c'}^2 \widetilde{\mathbf{w}}_{q \comma u \comma c}\HH \widetilde{\mathbf{\Phi}}_{v \comma c} \mathbf{G}_{v \comma c'}^* \mathbf{G}_{v \comma c'}\TT \widetilde{\mathbf{\Phi}}_{v \comma c}\HH \widetilde{\mathbf{w}}_{q \comma u \comma c}
	\label{eq:power_terms4}
\end{align}
\begin{align}
	&\widetilde{P}_{\text{z,} c} =
	\widetilde{\mathbf{w}}_{q \comma u \comma c}\HH \widetilde{\mathbf{K}}_{\text{RxA,} c} {\mathbf{R}}_{\text{z,} c} \widetilde{\mathbf{K}}_{\text{RxA,} c}\HH \widetilde{\mathbf{w}}_{q \comma u \comma c}
	\label{eq:power_terms5}
\end{align}
\begin{align}
	&\widetilde{P}_{\text{z,} c'} =
	\widetilde{\mathbf{w}}_{q \comma u \comma c}\HH \widetilde{\mathbf{K}}_{\text{RxB,} c} {\mathbf{R}}_{\text{z,} c'}^* \widetilde{\mathbf{K}}_{\text{RxB,} c}\HH \widetilde{\mathbf{w}}_{q \comma u \comma c}.
	\label{eq:power_terms6}
\end{align}
Note that $\sigma_{\text{x,} u \comma c}^2$ denotes the power of a single data stream of UE $u$ and thus the total power of the data streams of UE $u$ at subcarrier $c$ is equal to $Q_{u \comma c} \sigma_{\text{x,} u \comma c}^2$. 


\bibliographystyle{IEEEtran}
\bibliography{IEEEabrv,twc_2016_hakkarainen}

\begin{thebibliography}{10}
\providecommand{\url}[1]{#1}
\csname url@samestyle\endcsname
\providecommand{\newblock}{\relax}
\providecommand{\bibinfo}[2]{#2}
\providecommand{\BIBentrySTDinterwordspacing}{\spaceskip=0pt\relax}
\providecommand{\BIBentryALTinterwordstretchfactor}{4}
\providecommand{\BIBentryALTinterwordspacing}{\spaceskip=\fontdimen2\font plus
\BIBentryALTinterwordstretchfactor\fontdimen3\font minus
  \fontdimen4\font\relax}
\providecommand{\BIBforeignlanguage}[2]{{%
\expandafter\ifx\csname l@#1\endcsname\relax
\typeout{** WARNING: IEEEtran.bst: No hyphenation pattern has been}%
\typeout{** loaded for the language `#1'. Using the pattern for}%
\typeout{** the default language instead.}%
\else
\language=\csname l@#1\endcsname
\fi
#2}}
\providecommand{\BIBdecl}{\relax}
\BIBdecl

\bibitem{hakkarainen_precoded_2014}
A.~Hakkarainen, J.~Werner, K.~R. Dandekar, and M.~Valkama, ``Precoded massive
  {MU-MIMO} uplink transmission under transceiver {I/Q} imbalance,'' in
  \emph{Proc. {IEEE} {GLOBECOM} Workshops}, Dec. 2014, pp. 405--411.

\bibitem{hakkarainen_transceiver_2015}
A.~Hakkarainen, J.~Werner, M.~Renfors, K.~R. Dandekar, and M.~Valkama,
  ``Transceiver {I/Q} imbalance and widely-linear spatial processing in large
  antenna systems,'' in \emph{Proc. Int. Symp. on Wireless Commun. Syst.
  {(ISWCS)}}, Aug. 2015, pp. 651--655.

\bibitem{mirabbasi_classical_2000}
S.~Mirabbasi and K.~Martin, ``Classical and modern receiver architectures,''
  \emph{{IEEE} Commun. Mag.}, vol.~38, no.~11, pp. 132--139, Nov. 2000.

\bibitem{schenk_rf_2008}
T.~Schenk, \emph{{RF} Imperfections in High-rate Wireless Systems: Impact and
  Digital Compensation}, 1st~ed.\hskip 1em plus 0.5em minus 0.4em\relax
  Springer, 2008.

\bibitem{tarighat_compensation_2005}
A.~Tarighat, R.~Bagheri, and A.~Sayed, ``Compensation schemes and performance
  analysis of {IQ} imbalances in {OFDM} receivers,'' \emph{{IEEE} Trans. Signal
  Process.}, vol.~53, no.~8, pp. 3257--3268, Aug. 2005.

\bibitem{tarighat_joint_2007}
A.~Tarighat and A.~Sayed, ``Joint compensation of transmitter and receiver
  impairments in {OFDM} systems,'' \emph{{IEEE} Trans. Wireless Commun.},
  vol.~6, no.~1, pp. 240--247, Jan. 2007.

\bibitem{ozdemir_exact_2013}
{\"O}.~\"Ozdemir, R.~Hamila, and N.~Al-Dhahir, ``Exact {SINR} analysis of
  {OFDM} systems under joint {Tx/RX} {I/Q} imbalance,'' in \emph{Proc. {IEEE}
  {PIMRC}}, 2013, pp. 646--650.

\bibitem{ozdemir_exact_2014}
------, ``Exact average {OFDM} subcarrier {SINR} analysis under joint
  transmit-receive {I}/{Q} imbalance,'' \emph{IEEE Trans. Veh. Technol.},
  vol.~63, no.~8, pp. 4125--4130, Oct. 2014.

\bibitem{krone_capacity_2008}
S.~Krone and G.~Fettweis, ``On the {Capacity} of {OFDM} {Systems} with
  {Receiver} {I}/{Q} {Imbalance},'' in \emph{Proc. {IEEE} {ICC}}, May 2008, pp.
  1317--1321.

\bibitem{narasimhan_digital_2009}
B.~Narasimhan, D.~Wang, S.~Narayanan, H.~Minn, and N.~Al-Dhahir, ``Digital
  compensation of frequency-dependent joint {Tx}/{Rx} {I}/{Q} imbalance in
  {OFDM} systems under high mobility,'' \emph{IEEE J. Sel. Topics Signal
  Process.}, vol.~3, no.~3, pp. 405--417, jun 2009.

\bibitem{maham_impact_2012}
B.~Maham, O.~Tirkkonen, and A.~Hjorungnes, ``Impact of transceiver {I/Q}
  imbalance on transmit diversity of beamforming {OFDM} systems,'' \emph{{IEEE}
  Trans. Commun.}, vol.~60, no.~3, pp. 643--648, Mar. 2012.

\bibitem{ozdemir_sinr_2012}
{\"O}.~\"Ozdemir, R.~Hamila, and N.~Al-Dhahir, ``{SINR} analysis for
  beamforming {OFDM} systems under joint transmit-receive {I/Q} imbalance,'' in
  \emph{Proc. {IEEE} {PIMRC}}, 2012, pp. 1896--1901.

\bibitem{tarighat_mimo_2005}
A.~Tarighat and A.~Sayed, ``{MIMO} {OFDM} receivers for systems with {IQ}
  imbalances,'' \emph{{IEEE} Trans. Signal Process.}, vol.~53, no.~9, pp.
  3583--3596, Sep. 2005.

\bibitem{schenk_estimation_2006}
T.~C.~W. Schenk, P.~F.~M. Smulders, and E.~Fledderus, ``Estimation and
  compensation of frequency selective {TX/RX} {IQ} imbalance in {MIMO} {OFDM}
  systems,'' in \emph{Proc. {IEEE} {ICC}}, 2006, pp. 251--256.

\bibitem{schenk_performance_2007}
T.~C.~W. Schenk, E.~Fledderus, and P.~F.~M. Smulders, ``Performance impact of
  {IQ} mismatch in direct-conversion {MIMO} {OFDM} transceivers,'' in
  \emph{Proc. {IEEE} {RWS}}, 2007, pp. 329--332.

\bibitem{ozdemir_i/q_2013}
{\"O}.~\"Ozdemir, R.~Hamila, and N.~Al-Dhahir, ``{I/Q} imbalance in multiple
  beamforming {OFDM} transceivers: {SINR} analysis and digital baseband
  compensation,'' \emph{{IEEE} Trans. Commun.}, vol.~61, no.~5, pp. 1914--1925,
  May 2013.

\bibitem{qi_compensation_2010}
J.~Qi and S.~Aissa, ``Compensation for {HPA} nonlinearity and {I/Q} imbalance
  in {MIMO} beamforming systems,'' in \emph{Proc. {IEEE} {WiMob}}, 2010, pp.
  78--82.

\bibitem{tandur_joint_2007}
D.~Tandur and M.~Moonen, ``Joint adaptive compensation of transmitter and
  receiver {IQ} imbalance under carrier frequency offset in {OFDM}-based
  systems,'' \emph{{IEEE} Trans. Signal Process.}, vol.~55, no.~11, pp.
  5246--5252, Nov. 2007.

\bibitem{tubbax_compensation_2005}
J.~Tubbax, B.~Come, L.~Van~der Perre, S.~Donnay, M.~Engels, H.~De~Man, and
  M.~Moonen, ``Compensation of {IQ} imbalance and phase noise in {OFDM}
  systems,'' \emph{IEEE Trans. Wireless Commun.}, vol.~4, no.~3, pp. 872--877,
  May 2005.

\bibitem{zou_joint_2009}
Q.~Zou, A.~Tarighat, and A.~Sayed, ``Joint compensation of {IQ} imbalance and
  phase noise in {OFDM} wireless systems,'' \emph{IEEE Trans. Commun.},
  vol.~57, no.~2, pp. 404--414, Feb. 2009.

\bibitem{ishaque_capacity_2013}
A.~Ishaque, P.~Sakulkar, and G.~Ascheid, ``Capacity analysis of uplink
  multi-user {SC}-{FDMA} system with frequency-dependent {I}/{Q} imbalance,''
  in \emph{Proc. {Allerton}}, Oct. 2013, pp. 1067--1074.

\bibitem{yoshida_analysis_2009}
Y.~Yoshida, K.~Hayashi, H.~Sakai, and W.~Bocquet, ``Analysis and {Compensation}
  of {Transmitter} {IQ} {Imbalances} in {OFDMA} and {SC}-{FDMA} {Systems},''
  \emph{IEEE Trans. Signal Process.}, vol.~57, no.~8, pp. 3119--3129, Aug.
  2009.

\bibitem{picinbono_widely_1995}
B.~Picinbono and P.~Chevalier, ``Widely linear estimation with complex data,''
  \emph{{IEEE} Trans. Signal Process.}, vol.~43, no.~8, pp. 2030--2033, Aug.
  1995.

\bibitem{schreier_second-order_2003}
P.~Schreier and L.~Scharf, ``Second-order analysis of improper complex random
  vectors and processes,'' \emph{{IEEE} Trans. Signal Process.}, vol.~51,
  no.~3, pp. 714--725, Mar. 2003.

\bibitem{adali_complex-valued_2011}
T.~Adali, P.~Schreier, and L.~Scharf, ``Complex-valued signal processing: {The}
  proper way to deal with impropriety,'' \emph{{IEEE} Trans. Signal Process.},
  vol.~59, no.~11, pp. 5101--5125, Nov. 2011.

\bibitem{valkama_blind_2005}
M.~Valkama, M.~Renfors, and V.~Koivunen, ``Blind signal estimation in conjugate
  signal models with application to {I/Q} imbalance compensation,''
  \emph{{IEEE} Signal Process. Lett.}, vol.~12, no.~11, pp. 733--736, Nov.
  2005.

\bibitem{anttila_circularity-based_2008}
L.~Anttila, M.~Valkama, and M.~Renfors, ``Circularity-based {I/Q} imbalance
  compensation in wideband direct-conversion receivers,'' \emph{{IEEE} Trans.
  Veh. Technol.}, vol.~57, no.~4, pp. 2099 --2113, Jul. 2008.

\bibitem{gesbert_shifting_2007}
D.~Gesbert, M.~Kountouris, R.~Heath, C.-B. Chae, and T.~Salzer, ``Shifting the
  {MIMO} paradigm,'' \emph{{IEEE} Signal Process. Mag.}, vol.~24, no.~5, pp.
  36--46, Sep. 2007.

\bibitem{gomaa_multi-user_2011}
A.~Gomaa and N.~Al-Dhahir, ``Multi-user {SC}-{FDMA} systems under {IQ}
  imbalance: {EVM} and subcarrier mapping impact,'' in \emph{Proc. {IEEE}
  {GLOBECOM}}, Dec. 2011, pp. 1--5.

\bibitem{hakkarainen_interference_2014}
A.~Hakkarainen, J.~Werner, K.~Dandekar, and M.~Valkama, ``Interference
  suppression with antenna arrays in {OFDM} systems under transceiver {I/Q}
  imbalance,'' in \emph{Proc. {CROWNCOM}}, Jun. 2014, pp. 278--284.

\bibitem{hoydis_massive_2013}
J.~Hoydis, S.~ten Brink, and M.~Debbah, ``Massive {MIMO} in the {UL/DL} of
  cellular networks: How many antennas do we need?'' \emph{{IEEE} J. Sel. Areas
  Commun.}, vol.~31, no.~2, pp. 160--171, Feb. 2013.

\bibitem{larsson_massive_2014}
E.~Larsson, O.~Edfors, F.~Tufvesson, and T.~Marzetta, ``Massive {MIMO} for next
  generation wireless systems,'' \emph{{IEEE} Commun. Mag.}, vol.~52, no.~2,
  pp. 186--195, Feb. 2014.

\bibitem{lu_overview_2014}
L.~Lu, G.~Li, A.~Swindlehurst, A.~Ashikhmin, and R.~Zhang, ``An overview of
  massive {MIMO}: Benefits and challenges,'' \emph{IEEE J. Sel. Topics Signal
  Process.}, vol.~8, no.~5, Oct. 2014.

\bibitem{wimax_2012}
\BIBentryALTinterwordspacing
\emph{{IEEE Standard for WirelessMAN-Advanced Air Interface for Broadband
  Wireless Access Systems}}, {IEEE Standards Association} Standard
  802.16.1™-2012, sep 2012. [Online]. Available:
  \url{http://standards.ieee.org/getieee802/download/802.16.1-2012.pdf}
\BIBentrySTDinterwordspacing

\bibitem{ieee80211sg}
\BIBentryALTinterwordspacing
{Institute of Electrical and Electronics Engineers, Inc. ({IEEE})}. (2014)
  Status of {IEEE} 802.11 {HEW} study group. [Online]. Available:
  \url{http://www.ieee802.org/11/Reports/hew_update.htm}
\BIBentrySTDinterwordspacing

\bibitem{ieee80211tg}
\BIBentryALTinterwordspacing
{IEEE Standards Association}. (2014) {IEEE} 802.11 documents. [Online].
  Available: \url{https://mentor.ieee.org/802.11/documents}
\BIBentrySTDinterwordspacing

\bibitem{bjornson_massive_HW_2014}
E.~Bj\"ornson, M.~Matthaiou, and M.~Debbah, ``Massive {MIMO} systems with
  hardware-constrained base stations,'' in \emph{Proc. {IEEE} {ICASSP}}, May
  2014, pp. 3142--3146.

\bibitem{bjornson_massive_2014}
\BIBentryALTinterwordspacing
E.~Bj\"ornson, E.~G. Larsson, and M.~Debbah, ``Massive {MIMO} for maximal
  spectral efficiency: {How} many users and pilots should be allocated?''
  \emph{{IEEE} Trans. Wireless Commun.}, 2015. [Online]. Available:
  \url{http://arxiv.org/abs/1412.7102}
\BIBentrySTDinterwordspacing

\bibitem{anttila_frequency-selective_2008}
L.~Anttila, M.~Valkama, and M.~Renfors, ``Frequency-selective {I/Q} mismatch
  calibration of wideband direct-conversion transmitters,'' \emph{{IEEE} Trans.
  Circuits Syst. {II}, Exp. Briefs}, vol.~55, no.~4, pp. 359 --363, Apr. 2008.

\bibitem{luo_digital_2011}
F.-L. Luo, \emph{\BIBforeignlanguage{en}{Digital Front-End in Wireless
  Communications and Broadcasting: {Circuits} and Signal Processing}}.\hskip
  1em plus 0.5em minus 0.4em\relax Cambridge University Press, Sep. 2011.

\bibitem{litva_digital_1996}
J.~Litva, \emph{Digital Beamforming in Wireless Communications}.\hskip 1em plus
  0.5em minus 0.4em\relax Artech House Publishers, Aug. 1996.

\bibitem{gomaa_phase_2014}
A.~Gomaa and N.~Al-Dhahir, ``Phase {Noise} in {Asynchronous} {SC}-{FDMA}
  {Systems}: {Performance} {Analysis} and {Data}-{Aided} {Compensation},''
  \emph{IEEE Trans. Veh. Technol.}, vol.~63, no.~6, pp. 2642--2652, Jul. 2014.

\bibitem{trees_detection_2004}
H.~L.~V. Trees, \emph{\BIBforeignlanguage{en}{Detection, Estimation, and
  Modulation Theory, Optimum Array Processing}}.\hskip 1em plus 0.5em minus
  0.4em\relax John Wiley \& Sons, Apr. 2004.

\bibitem{3gpp_evolved_2014}
\emph{Evolved {Universal} {Terrestrial} {Radio} {Access} ({E}-{UTRA}); {User}
  {Equipment} ({UE}) radio transmission and reception}.\hskip 1em plus 0.5em
  minus 0.4em\relax The 3rd Generation Partnership Project (3GPP) Tech. Spec.
  V11.8.0, Release 11, TS36.101, Mar. 2014.

\bibitem{3gpp_evolved_2013}
\emph{Evolved {Universal} {Terrestrial} {Radio} {Access} ({E}-{UTRA}); {User}
  {Equipment} ({UE}) conformance specification; {Radio} transmission and
  reception}.\hskip 1em plus 0.5em minus 0.4em\relax The 3rd Generation
  Partnership Project (3GPP) Tech. Spec. V11.2.0, Release 11, TS36.521.-1, Sep.
  2013.

\bibitem{widrow_adaptive_1967}
B.~{Widrow et al.}, ``Adaptive antenna systems,'' \emph{Proc. {IEEE}}, vol.~55,
  no.~12, pp. 2143--2159, Dec. 1967.

\bibitem{johnson_modified_2007}
S.~Johnson and M.~Frigo, ``A modified split-radix {FFT} with fewer arithmetic
  operations,'' \emph{IEEE Trans. Signal Process.}, vol.~55, no.~1, pp.
  111--119, Jan. 2007.

\bibitem{haykin_adaptive_2002}
S.~Haykin, \emph{Adaptive Filter Theory}, 4th~ed.\hskip 1em plus 0.5em minus
  0.4em\relax Prentice Hall, 2002.

\bibitem{hakkarainen_widely-linear_2013}
A.~Hakkarainen, J.~Werner, K.~R. Dandekar, and M.~Valkama, ``Widely-linear
  beamforming and {RF} impairment suppression in massive antenna arrays,''
  \emph{J. Commun. and Netw.}, vol.~15, no.~4, pp. 383--397, Aug. 2013.

\bibitem{bjornson_circuit-aware_2014}
\BIBentryALTinterwordspacing
E.~Bj\"ornson, M.~Matthaiou, and M.~Debbah, ``Circuit-aware design of
  energy-efficient massive {MIMO} systems,'' in \emph{Proc. {ISCCSP}}, May
  2014. [Online]. Available: \url{http://arxiv.org/abs/1403.4851}
\BIBentrySTDinterwordspacing

\bibitem{bjornson_massive_2013}
E.~Bj\"ornson, J.~Hoydis, M.~Kountouris, and M.~Debbah, ``Massive {MIMO}
  systems with non-ideal hardware: {Energy} efficiency, estimation, and
  capacity limits,'' \emph{IEEE Trans. Inf. Theory}, vol.~60, no.~11, pp.
  7112--7139, Nov. 2014.

\bibitem{ngo_energy_2013}
H.~Q. Ngo, E.~Larsson, and T.~Marzetta, ``Energy and spectral efficiency of
  very large multiuser {MIMO} systems,'' \emph{IEEE Trans. Commun.}, vol.~61,
  no.~4, pp. 1436--1449, Apr. 2013.

\end{thebibliography}

\vspace*{-1\baselineskip}
\begin{IEEEbiography}[{\includegraphics[width=1in,height=1.25in,clip,keepaspectratio]{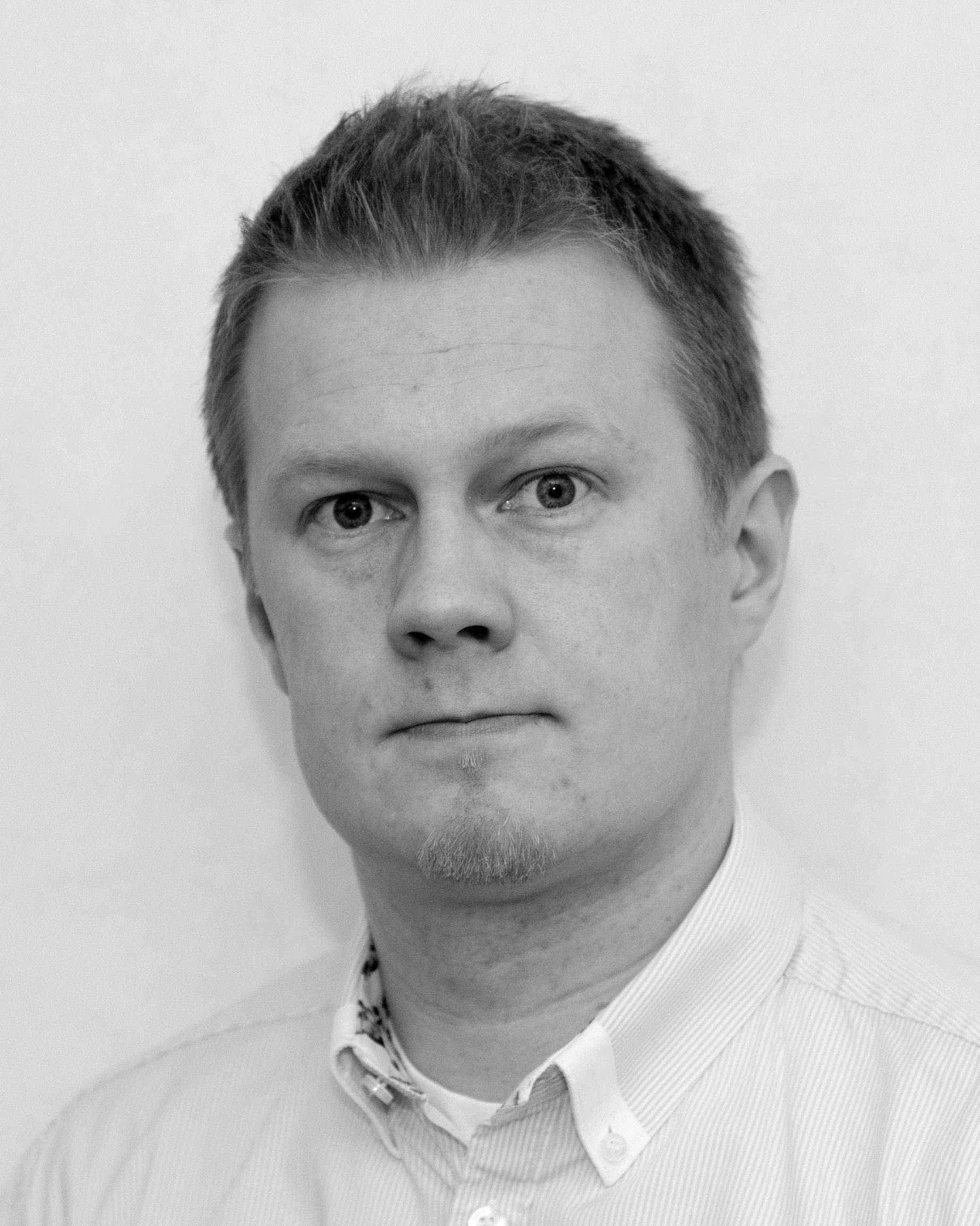}}]%
{Aki Hakkarainen}
(S'14) was born in Joensuu, Finland, in 1982. He received the M.Sc. (with honors) in communication electronics from Tampere University of Technology (TUT), Finland, in 2007. From 2007 to 2009, he was working as a RF design engineer with Nokia, Salo. From 2009 to 2011, he was working as a Radio system specialist with Elisa, Tampere. Currently he is working towards the Ph.D. degree at the Department of Electronics and Communications Engineering, TUT. His research interests include digital signal processing methods for flexible radio transceivers with a focus on RF impairment mitigation, and directional antennas in 5G networks with an emphasis on radio localization aspects.
\end{IEEEbiography}

\vspace*{-3\baselineskip}
\begin{IEEEbiography}[{\includegraphics[width=1in,height=1.25in,clip,keepaspectratio]{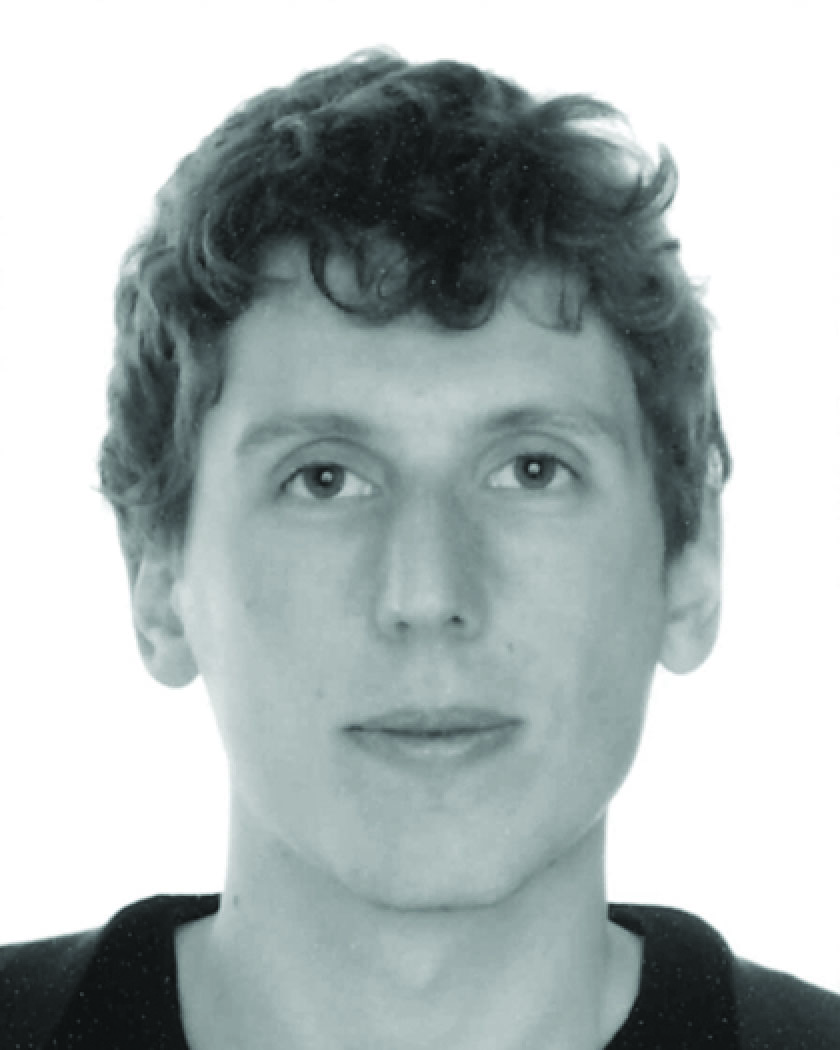}}]%
{Janis Werner}
was born in Berlin, Germany, in 1986. He received the Dipl.\ Ing.\ degree in electrical engineering from Dresden University of Technology (TUD), Germany in 2011 and his Ph.D.\ degree from Tampere University of Technology (TUT), Finland in 2015. His main research interests are localization with an emphasis on directional antenna-based systems as well as the utilization and further processing of location information in future generation mobile networks.
\end{IEEEbiography}

\vspace*{-3\baselineskip}
\begin{IEEEbiography}[{\includegraphics[width=1in,height=1.25in,clip,keepaspectratio]{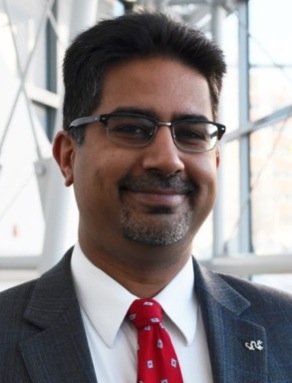}}]%
{Kapil R. Dandekar}
(S'95--M'01--SM'07) received the B.S. degree in electrical engineering from the University of Virginia in 1997.  He received the M.S. and Ph.D. degrees in Electrical and Computer Engineering from the University of Texas at Austin in 1998 and 2001, respectively. In 1992, he worked at the U.S. Naval Observatory and from 1993-1997, he worked at the U.S. Naval Research Laboratory. 

In 2001, Dandekar joined the Electrical and Computer Engineering Department at Drexel University in Philadelphia, Pennsylvania. He is currently a Professor in Electrical and Computer Engineering at Drexel University; the Director of the Drexel Wireless Systems Laboratory (DWSL); Associate Dean for Research and Graduate Studies in the Drexel University College of Engineering. DWSL has been supported by the U.S. National Science Foundation, Army CERDEC, National Security Agency, Office of Naval Research, and private industry.  Dandekar’s current research interests and publications involve wireless, ultrasonic, and optical communications, reconfigurable antennas, and smart textiles. Intellectual property from DWSL has been licensed by external companies for commercialization.   Dandekar is also a past member of the IEEE Educational Activities Board and co-founder of the EPICS-in-IEEE program. 
\end{IEEEbiography}

\begin{IEEEbiography}[{\includegraphics[width=1in,height=1.25in,clip,keepaspectratio]{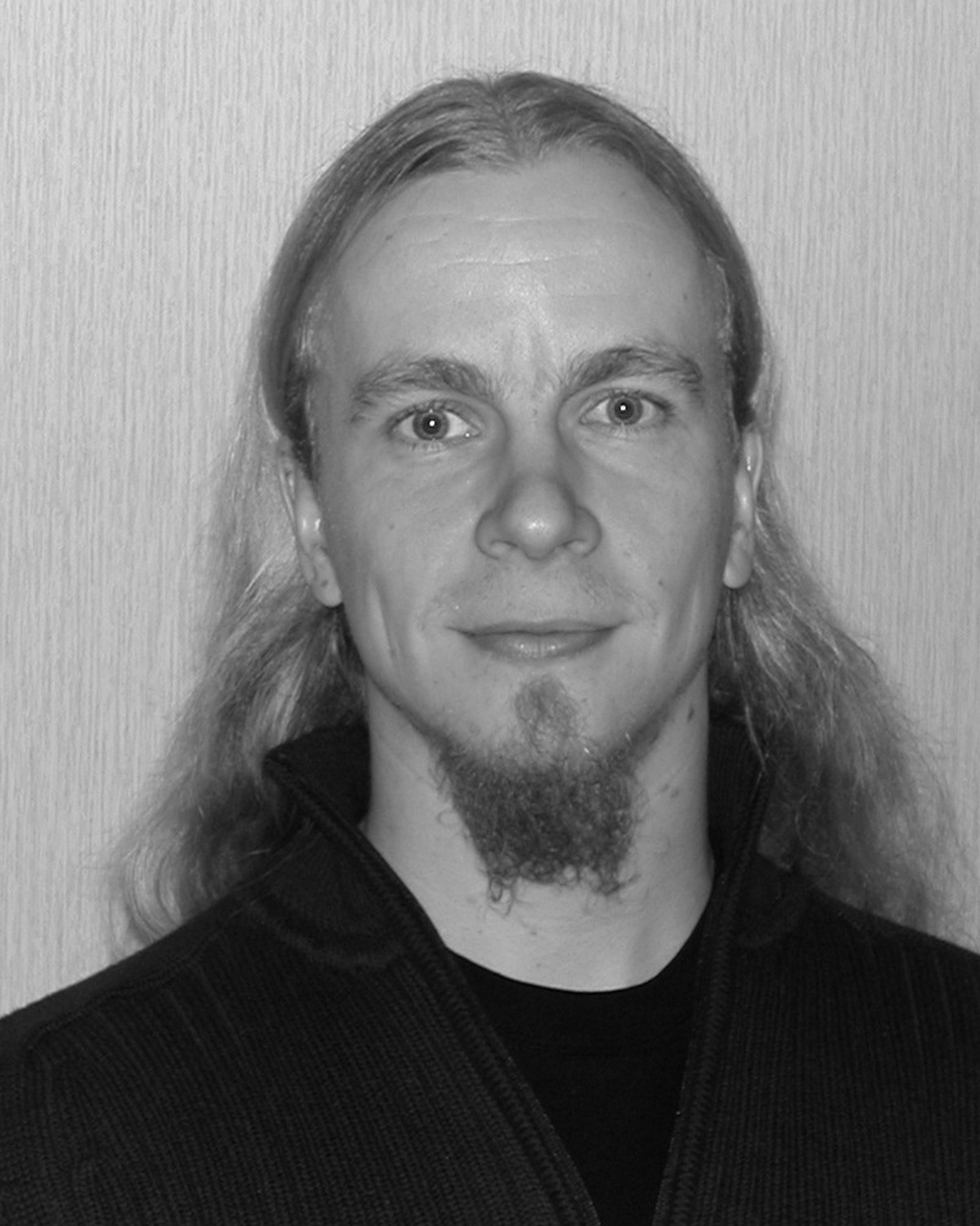}}]%
{Mikko Valkama}
was born in Pirkkala, Finland, on November 27, 1975. He received the M.Sc. and Ph.D. Degrees (both with honors) in electrical engineering (EE) from Tampere University of Technology (TUT), Finland, in 2000 and 2001, respectively. In 2002, he received the Best Ph.D. Thesis -award by the Finnish Academy of Science and Letters for his dissertation entitled "Advanced I/Q signal processing for wideband receivers: Models and algorithms". In 2003, he was working as a visiting researcher with the Communications Systems and Signal Processing Institute at SDSU, San Diego, CA. Currently, he is a Full Professor and Department Vice-Head at the Department of Electronics and Communications Engineering at TUT, Finland. His general research interests include communications signal processing, estimation and detection techniques, signal processing algorithms for software defined flexible radios, cognitive radio, full-duplex radio, radio localization, 5G mobile cellular radio, digital transmission techniques such as different variants of multicarrier modulation methods and OFDM, and radio resource management for ad-hoc and mobile networks.
\end{IEEEbiography}

\end{document}